\documentclass[preprint,aip,cha]{revtex4-1}

\usepackage[english]{babel}
\usepackage{float}
\usepackage{lineno}
\usepackage{graphicx}
\usepackage{dcolumn}
\usepackage{bm}
\usepackage{amsmath}
\usepackage{amssymb}
\usepackage{physics}
\usepackage{datetime}
\usepackage{hyperref}
\usepackage{bigints}
\usepackage{siunitx}
\usepackage{subcaption}
\usepackage{caption}
 \usepackage{todonotes}
\usepackage{color, colortbl}
\usepackage{cleveref}
\usepackage{multirow}
\usepackage{diagbox}
\usepackage{rotating}
\usepackage[normalem]{ulem}
\definecolor{Gray}{gray}{0.9}
\begin{document}

\title{Mode transitions and spoke structures in $\mathbf{E} \cross \mathbf{B}$ Penning discharge}
\author{M. Tyushev\footnote{Electronic mail: mikhail.tyushev@usask.ca; Corresponding author}}
\author{\firstname{M.}~\surname{Papahn Zadeh}}
\affiliation{Department of Physics and Engineering Physics, University of Saskatchewan, Saskatoon SK S7N 5E2, Canada}
\author{\firstname{N. S.}~\surname{Chopra}}
\author{\firstname{Y.}~\surname{Raitses}}
\author{\firstname{I.}~\surname{Romadanov}}
\affiliation{Princeton Plasma Physics Laboratory, Princeton, New Jersey 08540, USA}
\author{\firstname{A.}~\surname{Likhanskii}}
\affiliation{Applied Materials Inc, 35 Dory Rd, Gloucester, MA 01930, USA }
\author{\firstname{G.}~\surname{Fubiani}}
\author{\firstname{L.}~\surname{Garrigues}}
\affiliation{LAPLACE, Université de Toulouse, CNRS, INPT, UPS, 118 Route de Narbonne, 31062 Toulouse, France}
\author{\firstname{R.}~\surname{Groenewald}}
\affiliation{TAE Technologies Inc, 19631 Pauling, Foothill Ranch, CA 92610, USA}
\author{\firstname{A.}~\surname{Smolyakov}}
\affiliation{Department of Physics and Engineering Physics, University of Saskatchewan, Saskatoon SK S7N 5E2, Canada}

%\renewcommand\linenumberfont{\normalfont\scriptsize}

%\linenumbers 

\begin{abstract}
   Two-dimensional particle-in-cell simulations in the {(radial-azimuthal)} plane perpendicular to the  axial direction of a cylindrical $\mathbf{E}\cross \mathbf{B}$ Penning discharge are presented. The low-pressure discharge is self-consistently supported by plasma ionization from the electron beam injected axially, along the direction of the external magnetic field. It is shown that with the increasing strength of the external magnetic field, the discharge undergoes a sequence of transitions between several azimuthal modes. Azimuthal $m>1$ spiral arm structures are excited at low magnetic field values as plasma confinement improves and the radial density profile becomes peaked. With a larger field, spiral arms with $m>1$ are replaced by the $m=1$ spoke mode, most clearly seen in plasma density. A transition from spiral arms to the spoke regime occurs when the plasma potential in the center changes from weakly positive (or zero) to negative. Further increase of the magnetic field results in a well-developed $m=1$ spoke mode with additional small-scale higher frequency $m>1$ structures inside and around the spoke. It is shown that while ionization and collisions affect some characteristics of the observed fluctuations, the basic features of the spoke and $m>1$ spiral structure remained similar without ionization. The role of energy conservation in small-scale high-frequency modes and spoke dynamics is discussed. It is demonstrated that in regimes with the $m=1$ spoke mode, additional $m=4$ harmonics of the ion and electron fluxes to the wall appear due to the square boundary.  The frequency of the $m=1$ mode is weakly affected by the geometry of the boundary.  
   
   % simulations performed with WarpX PIC code utilizing the energy-conserving Yee grid. It is shown that switching to the collocated grid results in the $30\% $ energy error and  

%   \textcolor{purple}{The latter regime occurs when the injected axial beam energy contribution to ionization is mainly due to the kinetic energy of the beam rather than self-induced radial electric field and current} %The scaling of the $m=1$ spoke rotation frequency with magnetic field and mass/species scalings confirm the gradient-drift (lower-hybrid and Simon-Hoh)  nature of the structures.   
\end{abstract}

\maketitle

%----------------------------------------------
%---------------- INTRODUCTION ----------------
%----------------------------------------------

\section{Introduction}

Magnetic fields improve plasma confinement and are often employed to increase plasma density in low-temperature plasma devices. In typical applications with a moderate magnetic field, the electrons are fully magnetized and confined by the magnetic field while ions are only weakly affected by the magnetic field and can be controlled by the electric field.  Instabilities that occur in such partially magnetized plasmas due to plasma inhomogeneities and electric fields have long been a subject of active studies, e.g. see Refs. \onlinecite{KaganovichPoP2020} and \onlinecite{BoeufPoP2023} and references therein. Linear
theory \cite{SmolyakovPPCF2017,BoeufPoP2023} predicts a variety of instabilities that may occur in partially magnetized plasmas.  However,  nonlinear regimes and saturation of the instabilities have to be studied with numerical simulations.

A straight cylindrical configuration with an axial magnetic field and radial electric field (the so-called $\mathbf{E}\cross\mathbf{B}$ configuration) is a prototype for the Penning discharges used in many technological applications \cite{AbolmasovPSST2012}.  One of the intriguing and important phenomena that dominate the radial plasma transport is a large-scale, $m=1$ azimuthally rotating
structure, the so-called spoke,  typically observed as a large scale plasma density perturbation. Such spoke-type structures and multiple helicity modes in partially magnetized $\mathbf{E}\cross\mathbf{B}$ plasmas were
 observed experimentally in Penning and other  $\mathbf{E}\cross\mathbf{B}$ plasma systems \cite{SakawaPFB1993,SakawaPRL1992,RodriguezPoP2019,EllisonPoP2012,McDonaldIEEE2011,HecimovicPSST2015,PrzybockiPRL2024,JunePoP2023,JunePSST2023} and also  studied in numerical simulations under various conditions \cite{CarlssonPoP2018,PowisPoP2018,LiangPSST2021,BoeufFP2014,MatyasPSST2019s}.  
 
Here, we study the spoke and spiral arms structures in  a 2D radial-azimuthal geometry using Particle-in-Cell
(PIC) WarpX code\cite{WarpX}. We study the square as well as the circular geometry of the external boundary.  In both cases, Cartesian mesh is used thus avoiding the issues related to the region near $r=0$ in the cylindrical coordinates. The code is run on the Mist server, \url{https://docs.scinet.utoronto.ca/index.php/Mist}, of the Digital Research Alliance of Canada with one GPU of NVIDIA V100-SMX2-32GB type per one separate simulation.  The general setup of simulations is similar to that of  Ref.\onlinecite{TyushevPoP2023}.  Here we focus on the conditions for the spoke and spiral arms formations, the role of ionization, and consider the low pressure discharge of 0.44 mTorr while the higher pressure of 40 mTorr was considered in Ref.  \onlinecite{TyushevPoP2023}. 
    In this paper, we propose the mechanism for the spiral arms formation, show the coexistence of the $m=1$ spoke and $m>1$ small-scale high-frequency modes, and demonstrate that geometrical effects of the square boundary result in the modulation of the electron and ion fluxes to the wall at the fourth harmonics of the spoke frequency. To clarify the role of the ionization processes in the mechanism of the spoke, we have performed the simulations by turning off all collisional processes (including the ionization) and replacing the ionization with equivalent electron and ion sources. In this study, we observe the formation of the $m=1$ spoke with roughly similar characteristics, thus demonstrating that the mechanism of the spoke instability is not related to the ionization processes.  

In \Cref{sec:SimModel}, the PIC simulation model, and the parameters of
simulations are described in detail. In \Cref{sec:AzimuthalStr}, we
investigate the role of the electric field in the structure excitation and demonstrate the transitions between the $m>1$ spiral arms and the $m=1$ spoke regimes. The role of energy conservation in the spoke formation is
discussed in \Cref{sec:RoleOfEnergyCons}. In \Cref{sec:100mA}, we show the coexisting $m=1$ spoke and multiple $m>1$ high-frequency modes.  The effect of ionization on the spoke formation is discussed in \Cref{sec:NoIoniz}. In \Cref{sec:EffectOfGeometry}, we examine the effects of geometry on the spoke dynamics.  \Cref{sec:SummaryDiscussion} provides a summary and discussion of the results.

% Additionally, in our previous study (Ref.  \onlinecite{TyushevPoP2023}) we did not observe any changes after transfer to cylindrical coordinates

%---------------------------------------------
%---------------- MODELS ----------------
%---------------------------------------------
\section{The simulation model}\label{sec:SimModel}

%$\nu_{\varepsilon} = \SI{0.95e7}{s^{-1}}$)
\begin{figure}[htp]
\centering
\captionsetup{justification=raggedright,singlelinecheck=false}
\includegraphics[width=0.5\textwidth]{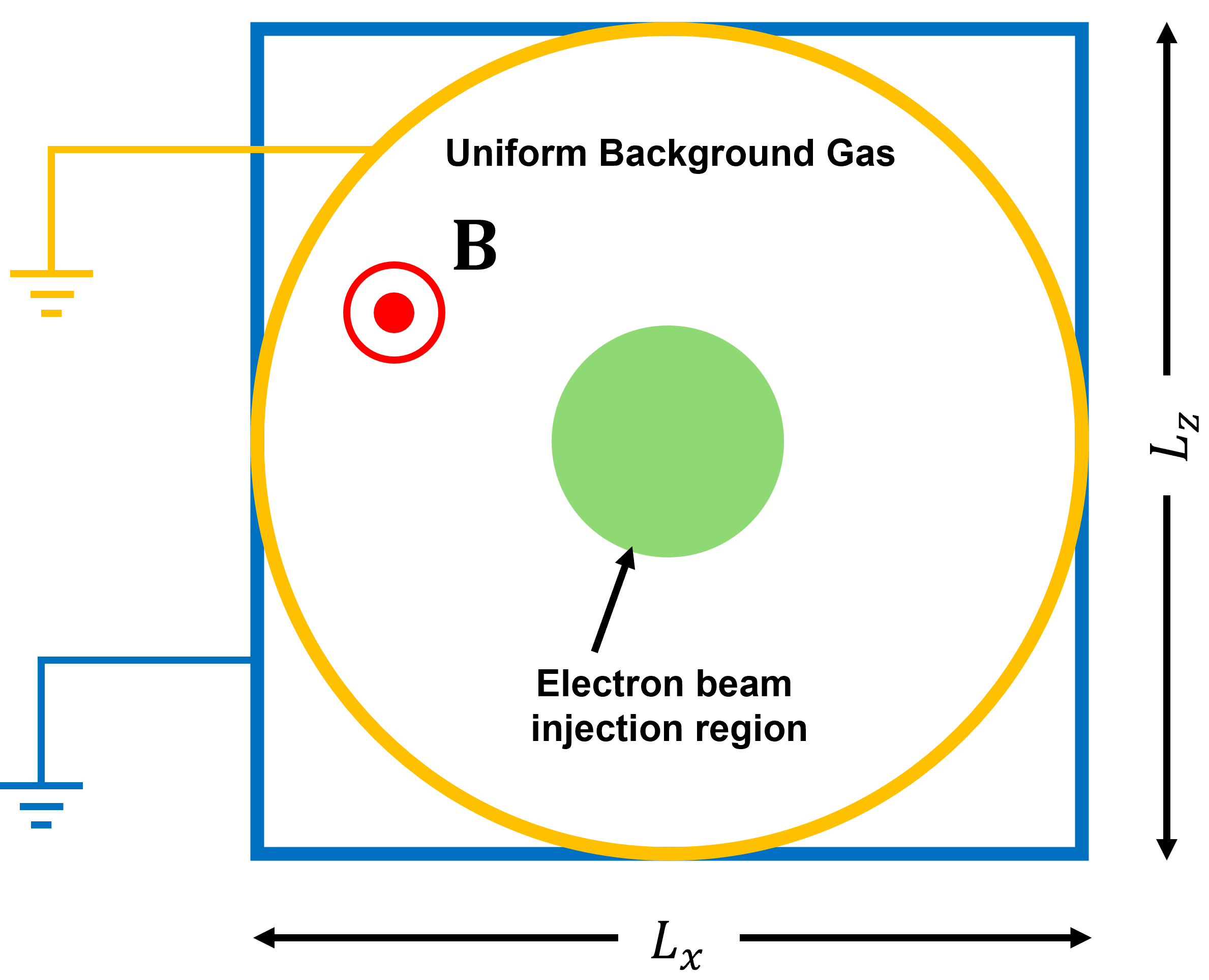}
\caption{The cross-section of the modeling region. The base case simulations are performed with a square boundary. \Cref{sec:EffectOfGeometry} simulations are done with a circular boundary. The potential of the boundary is zero in all cases.}
\label{fig:SimulationSetUp}
\end{figure}

The setup of our numerical experiments aims to reproduce the conditions similar to experimental conditions for Penning discharges and ion plasma sources \cite{RodriguezPoP2019,ChopraJAP2024,JunePoP2023,JunePSST2023}. In our simulations, plasma discharge is supported by the ionization from the energetic electrons injected along the magnetic field in the y-direction (outward of the page in Fig. \ref{fig:SimulationSetUp}). The electron beam is
uniform and injected into a circular region of a radius $R=1.5\;\text{cm}$, Fig. \ref{fig:SimulationSetUp}. In addition to one-step electron impact ionization $Ar + e \longrightarrow Ar + e + e$, for electrons with the energy above $15.7 \;\text{eV}$, excitation reactions $Ar + e \longrightarrow Ar^* + e$, with the excitation energy of  $11.55\;\text{eV}$,  and elastic collisions are included.
The electron-neutral isotropic scattering cross-sections included in WarpX distribution were sourced from Ref. \onlinecite{zatsarinny2004b}, excitation cross-sections from Ref.  \onlinecite{yamabe1983measurement},  and ionization cross-sections from Ref.  \onlinecite{bordage2013comparisons}. The beam electrons are injected along the magnetic field in the direction opposite to the field with $E_b=30\;\text{eV}$ axial energy and Maxwellian $T_e=5\;\text{eV}$ temperature throughout the whole simulation duration. The background neutral gas is maintained at a constant density throughout the entire simulation. The injected electron beam is represented by macroparticles introduced at each simulation time step, corresponding to the total beam current of $1\;\text{mA}$. Technically, in two-dimensional geometry, which neglects the third direction, the injected current along the unresolved direction should be measured in units of the current per unit of length, A/m. This dimension however can also be absorbed directly into the cell volume \cite{Brieda}, allowing the units of the injected current to remain in Amperes. The parameters of the simulations for the base case are given in \Cref{table:PhysicParamBaseCas}.

The 2D simulations are performed on a uniform Cartesian grid in the $x-z$ plane, with particle velocities in 3D $v_x-v_y-v_z$ space. A uniform axial magnetic field in the $y$ direction is applied perpendicular to the simulation domain, all incoming particles are absorbed at the outer boundary with Dirichlet conditions for the potential, $\Phi=0$. 
 In the base case simulations, we use the rectangular geometry for the discharge boundary, as shown in \Cref{fig:SimulationSetUp}. In the \Cref{sec:EffectOfGeometry}, we examine the effect of a circular grounded boundary.  

The simulations domain spans $L_x=L_z=10\;\text{cm}$ and is discretized into a grid of $n_x \times n_z=256\times256$ cells in the $x-$ and $z-$ directions, resulting in a mesh resolution of $\Delta x=0.39\;\text{mm}$. This resolution adequately resolves the Debye length through the simulations, as confirmed by direct calculations of electron temperature and plasma concentration in each cell. For the parameters of our simulations, at the center of the domain with, $B=100\;$ G, we obtain the density of $n=10^{15} \;\text{m}^{-3}$, and temperature of $T_e=5.5\;$eV. This results in a ratio of Debye length to mesh resolution of $\lambda_D/\Delta x = 1.41 $. Similarly, the minimum ratio for a magnetic field of $B=50\;$G is $2$, and for $B=220\;$ G the Debye length is resolved by a factor of $4$. 

The simulation time step is $\Delta t=50\;\text{ps}$ that resolves the electron cyclotron period and local plasma oscillation period satisfying the practical criteria of $\omega_{pe,ce} \Delta t<0.2$\cite{BirdsallBook}.
Electron cyclotron frequency for simulation with magnetic field of $B=100\;\text{G}$ is $\omega_{ce}=1.75\times10^9\;\text{rad/s}$, and the characteristic value of plasma frequency is $\omega_{pe}= 1.78\times10^9\;\text{rad/s}$ for $n_p = 10^{15}\;\text{m}^{-3}$. The spatial grid and time step are chosen to ensure that the Courant-Friedrichs-Lewy (CFL) condition is comfortably met. For the fastest particles, whose velocity is approximately equal to $v \approx 3v_{the}=4\times 10^6 \;\text{m/s}$ for temperature $T_e=5\;$eV we have: $\text{CFL} =  \sqrt{2} v \Delta t / \Delta x < 1$. The :q
ght of a single macroparticle was equivalent to $312500$ of real particles for simulations with Argon gas, both for electrons and ions.   This value results in an average of $145$ particles per cell (PPC). In simulations, the magnetic field is included for electrons and ions.  Plasma remains quasineutral to a high degree on average and for perturbations, as seen from the comparison of electron and ion densities in Fig. \ref{fig:DensSnap}.   

In simulations, injected electrons' energy exceeds Argon atoms' ionization energy and most of the energy for ionization is provided by the electron beam. Additional energy can be deposited from the radial current and potential drop across the radial direction\cite{TyushevPoP2023}.

\begin{table}[htp]
\centering
\caption{Physical parameters for the base case simulations with Argon.}
\label{table:PhysicParamBaseCas}
\begin{tabular}{||p{5cm} p{3cm} p{3cm}||}
\hline\hline 
\textbf{Property} & \textbf{Symbol} & \textbf{Value}  \\ \hline\hline
Magnetic Field & $B(\text{G})$ & $10$, $50$, $100$, $220$ \\
Electron beam energy & $E_{b}(\text{eV})$ & $30$ \\ 
Electron beam current & $I_e(\text{A})$ &  $0.001$ \\ 
Neutral temperature & $T_n(\text{K})$ & $300$   \\ 
Neutral pressure & $P_n(\text{mTorr})$ & $0.44$ \\  
Neutral density & $n_n(\text{m}^{-3})$ & $1.4\times 10^{19}$  \\ \hline \hline
\end{tabular}
\end{table}

Our study does not consider axial losses and ion collisions, thus all ions created by ionization reach the grounded electrode radially. In the actual 3D geometry, a fraction of ions created by ionization is lost axially. Since our simulations do not include the y-direction, such axial losses are not accounted for. Assuming that all ions are collected radially by the grounded electrode is a reasonable approximation for a discharge that is very long in the axial direction, since the axial losses scale as $R/L<1$, where $R$ and $L$ are the characteristic dimensions in radial and axial directions, respectively.   Although axial losses could theoretically be simulated in 2D geometry by removing ions at a certain rate, this was not done in this study. At $0.44\;$mTorr, the ion mean free path is large and ions can be assumed collisionless. 
Investigating these effects and developing methods to represent boundary conditions along the y-axis in a 2D model are left for future work.

\section{The azimuthal structures and mode transitions}\label{sec:AzimuthalStr}
\begin{figure}[hbp]
\centering
\captionsetup[subfigure]{labelformat=empty}  %\captionsetup[subfigure]{}
\subcaptionbox{\label{fig:DensSnap-10g}}{\includegraphics[width=0.78\linewidth]{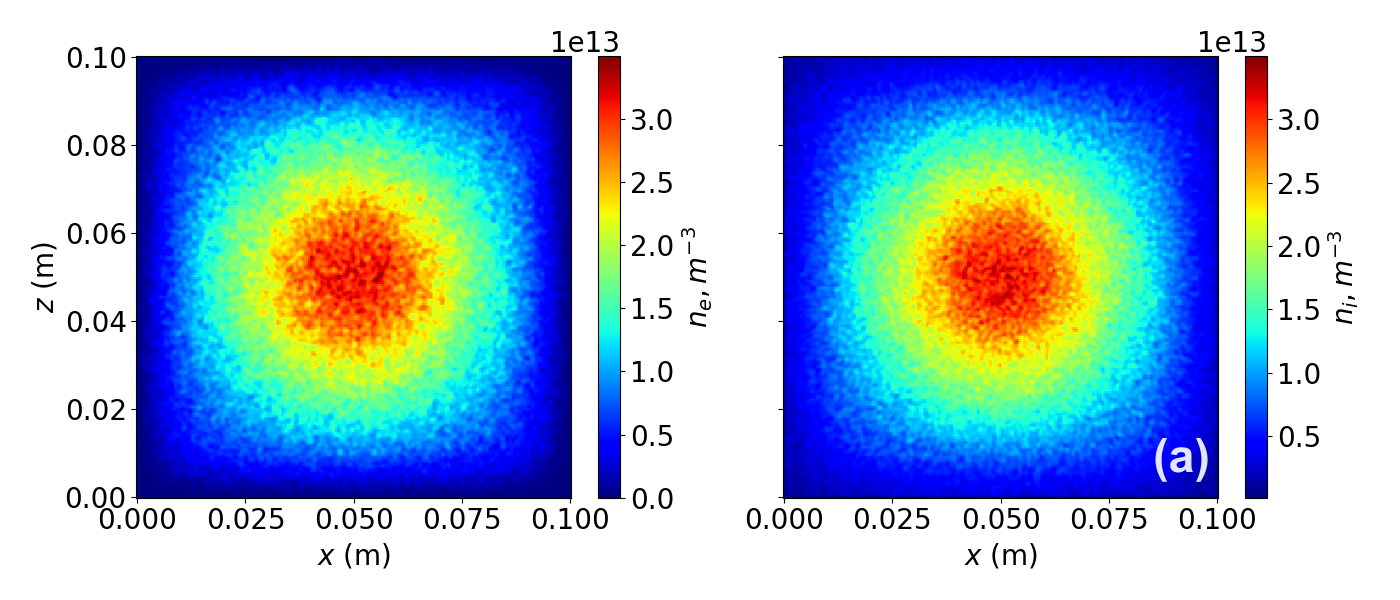}}\hfill
\vspace{-1.4cm}
\subcaptionbox{\label{fig:DensSnap-50g}}{\includegraphics[width=0.78\linewidth]{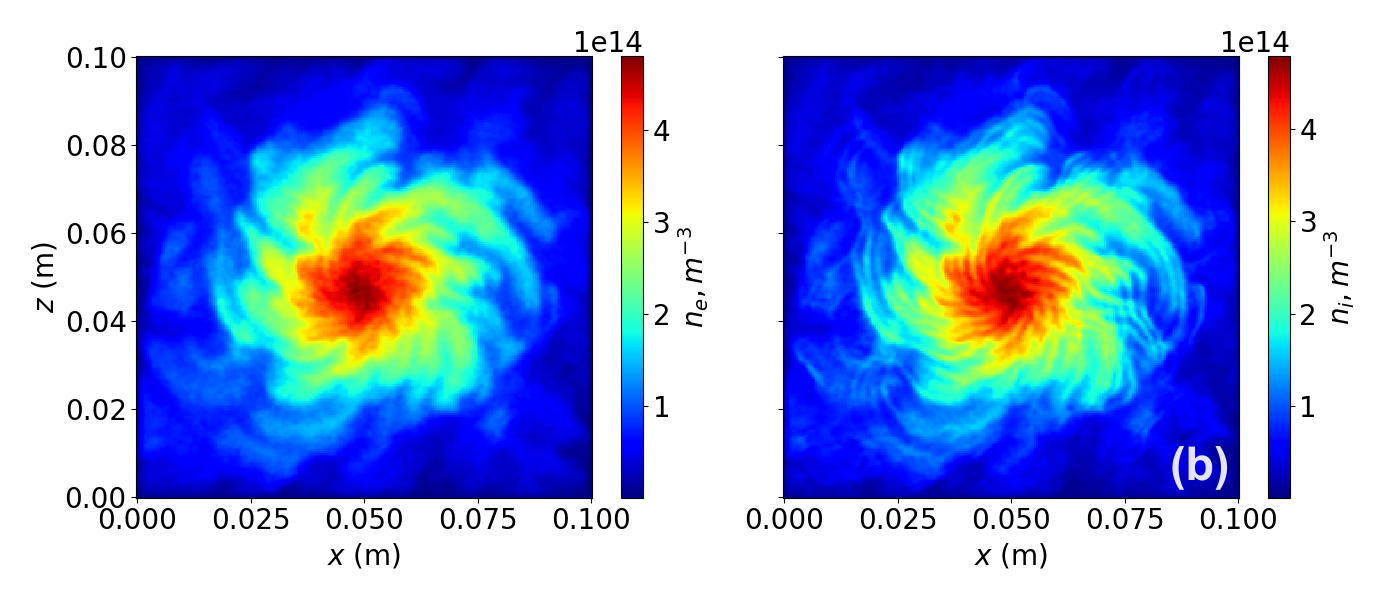}}\hfill
\vspace{-1.4cm}
\subcaptionbox{\label{fig:DensSnap-100g}}{\includegraphics[width=0.78\linewidth]{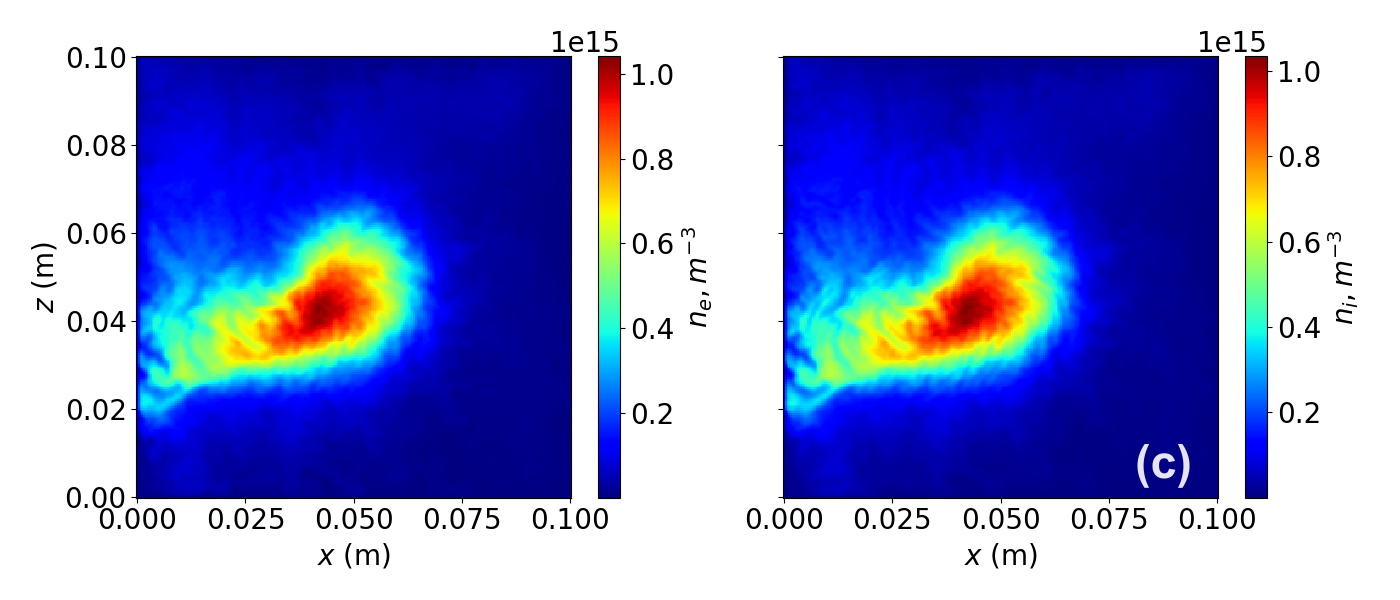}}\hfill
\vspace{-1.4cm}
\subcaptionbox{\label{fig:DensSnap-220g}}{\includegraphics[width=0.78\linewidth]{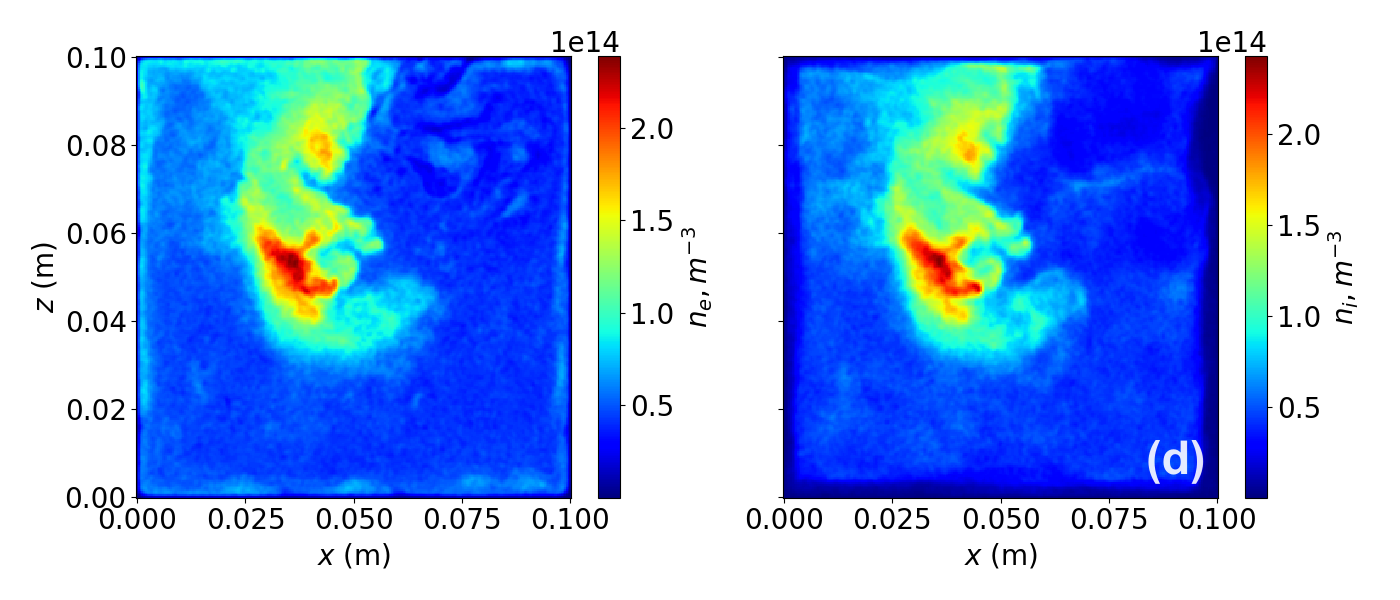}}\hfill
\vspace{-1.4cm}
 \captionsetup{justification=raggedright,singlelinecheck=false}
\caption{Snapshots of the electron (left) and ion (right) densities for the different magnitudes of the magnetic field: (a) $B=10\;$ G at $6.71\times 10^{-4}\;$s; (b) $B=50\;$ G at $8.86\times 10^{-4}\;$s (\href{https://www.youtube.com/watch?v=AyuNx3beVps}{YouTube video}); (c) $B=100\;$ G at $2.96\times 10^{-3}\;$s (\href{https://www.youtube.com/watch?v=Mf8vJmAvsQM}{YouTube video}); (d) $B=220\;$ G at $1.27\times 10^{-3}\;$ s (\href{https://www.youtube.com/watch?v=RlQdVlMbHGI}{YouTube video}).}
\label{fig:DensSnap}
\end{figure}

\begin{figure}[htp]
\centering
\captionsetup[subfigure]{}
\subcaptionbox{\label{fig:RadialProfi_ni}}{\includegraphics[width=0.49\linewidth]{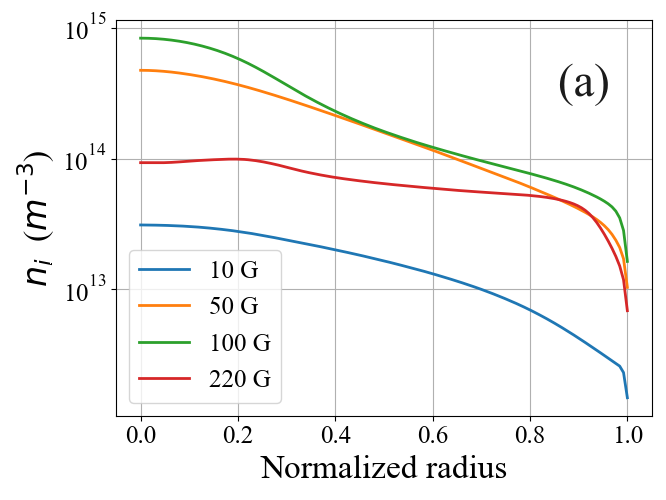}}
\subcaptionbox{\label{fig:RadiaProfi_phi}}{\includegraphics[width=0.49\linewidth]{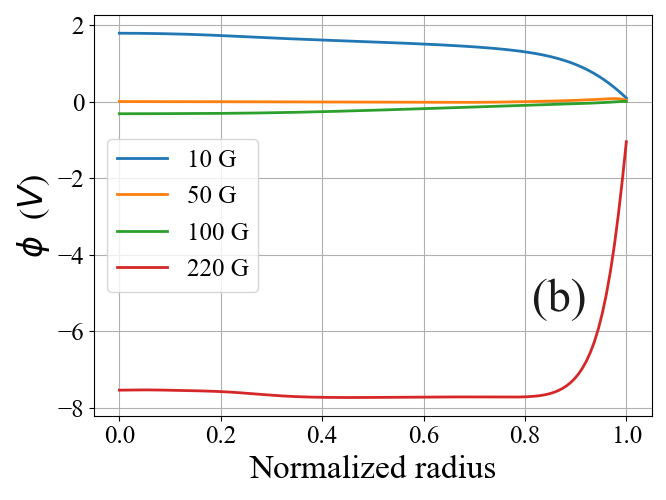}}
\vspace{-1.4cm}
\subcaptionbox{\label{fig:RadialProfi_Er}}{\includegraphics[width=0.49\linewidth]{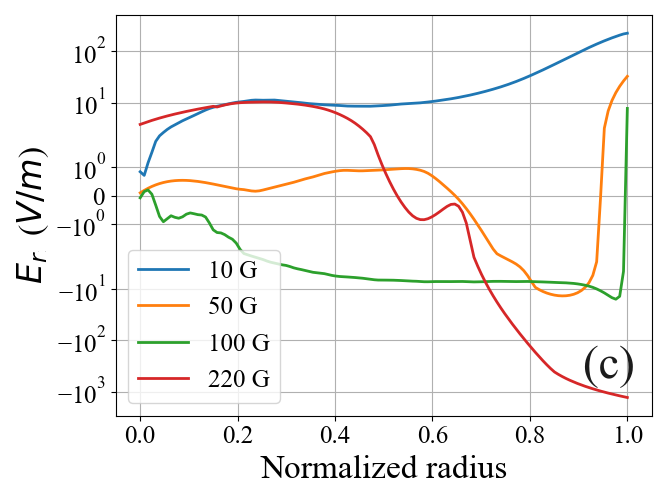}} 
\subcaptionbox{\label{fig:RadialProfi_Te}}{\includegraphics[width=0.49\linewidth]{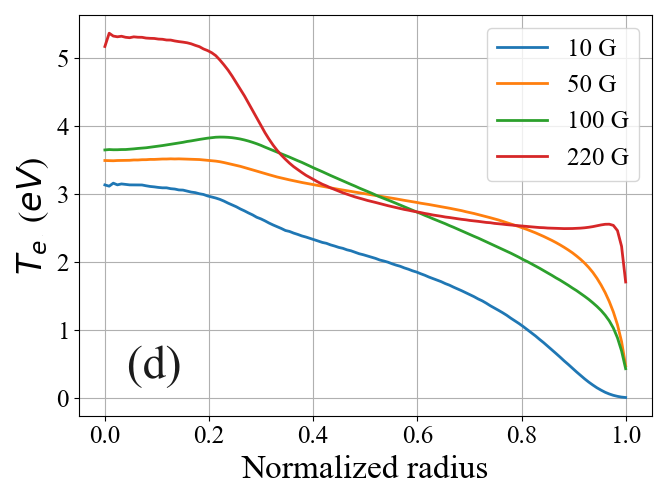}}
\captionsetup{justification=raggedright,singlelinecheck=false}
\caption{Averaged radial profiles of plasma parameters for different magnitudes of the magnetic field, the radius from the center of the domain is normalized to the largest radius which is $5 \;$cm, so the normalized injected radius is located at $0.3$: (a) ion density; (b) potential; (c) electric field; (d) electron temperature.}\label{fig:RadialProfiles}
\end{figure}

In this section, we demonstrate the sequence of transitions in the structure of the azimuthal modes occurring in the discharge as the magnetic field is increased.  Figure \ref{fig:DensSnap} displays a sequence of 2D maps representing the electron and ion density for four values of the magnetic field from $10\;$G (the first row) to $220\;$G (the last row). 
\Cref{fig:RadialProfiles} shows the radial profiles of plasma parameters for the same values of the magnetic field. Note that while Fig. \ref{fig:DensSnap} shows instantaneous values, Fig. \ref{fig:RadialProfiles} shows time-averaged radial plasma parameters in the saturated state. The time window averaging for $B=10\;$G and $B=50\;$G is the last $500\; \mu$s. For $B=100\;$G, it is the last $1960\;\mu$s, equivalent to $6$ spoke cycles. For $B=220\;$G, it is the last $2000\;\mu$s, equivalent to $50$ spoke cycles. 

At the lowest magnetic field intensity of $B=10\;$G, one observes mostly uniform (symmetric) plasma density with little signs of any structures. For the larger magnetic field of $B=50\;$G, we observe the excitation of the azimuthal $m>1$ spiral arm structures, Fig. \ref{fig:DensSnap}b. Similar structures were also observed in simulations of Ref. \onlinecite{LuckenPoP2019}.   Further increase of the magnetic field to $B=100\;$G results in a large scale $m=1$ coherently counterclockwise rotating structure along with $\mathbf{E}\cross\mathbf{B}$ drift, the so-called spoke. The $m=1$ slowly rotating structure coexists with fast spiral $m>1$ modes that occur at the edges and inside of the $m=1$ mode. Further increase of the magnetic field to $B=220\;$G leads to a more intense and violent spoke structure which also shows additional structures (blobs) that break away and move radially from the rotating spoke. 
One can note in  Fig.  \ref{fig:RadialProfiles}a that with increasing magnetic field plasma confinement is improving resulting in higher plasma density. This trend persists from $B=10\;$G to $B=100\;$G. However, one observes that for $B=220\;$G, the average plasma density drops again due to the large radial transport from the intense spoke and radially moving density patches/blobs. 

\begin{figure}[htp]
\centering
\captionsetup[subfigure]{labelformat=empty}
\subcaptionbox{\label{fig:Current-10g}}{\includegraphics[width=0.8\linewidth]{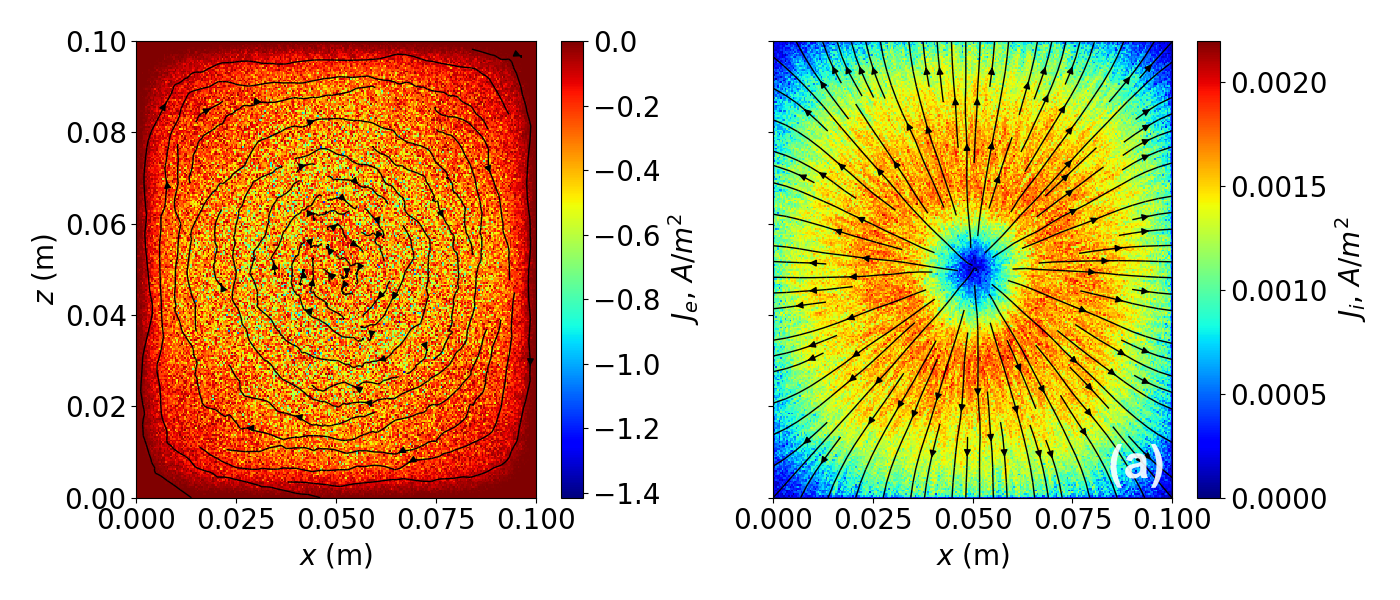}}\hfill
\vspace{-1.4cm}
\subcaptionbox{\label{fig:current-50g}}{\includegraphics[width=0.8\linewidth]{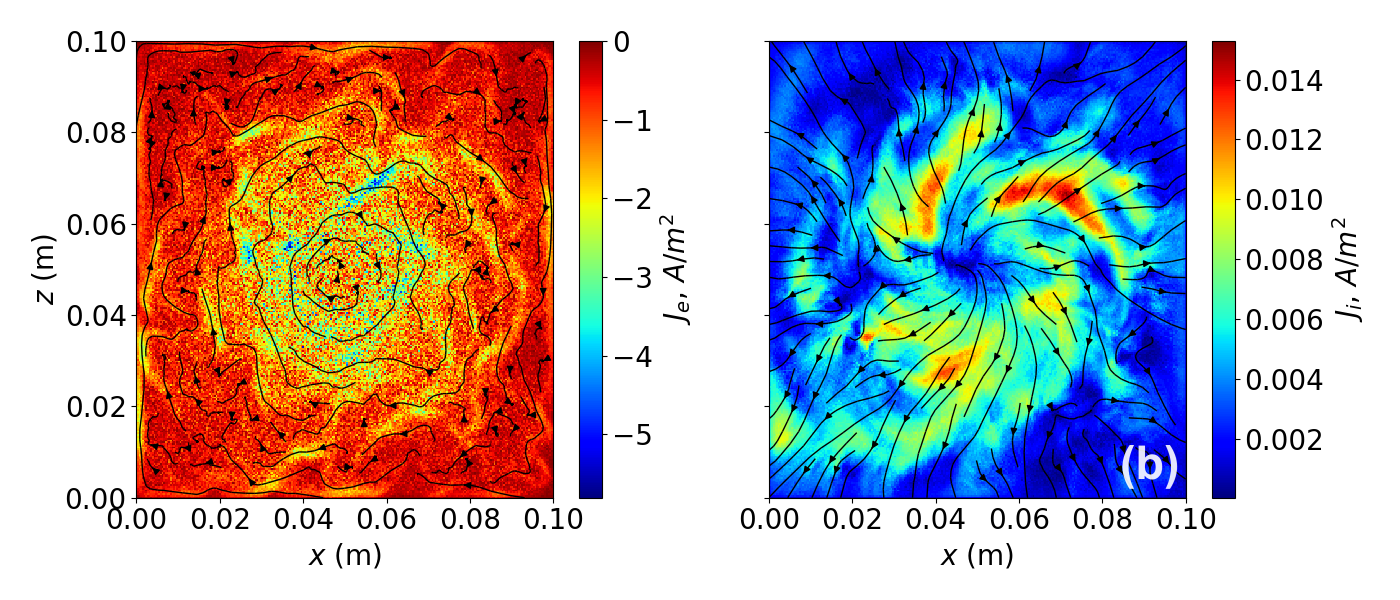}}\hfill
\vspace{-1.4cm}
\subcaptionbox{\label{fig:current-100g}}{\includegraphics[width=0.8\linewidth]{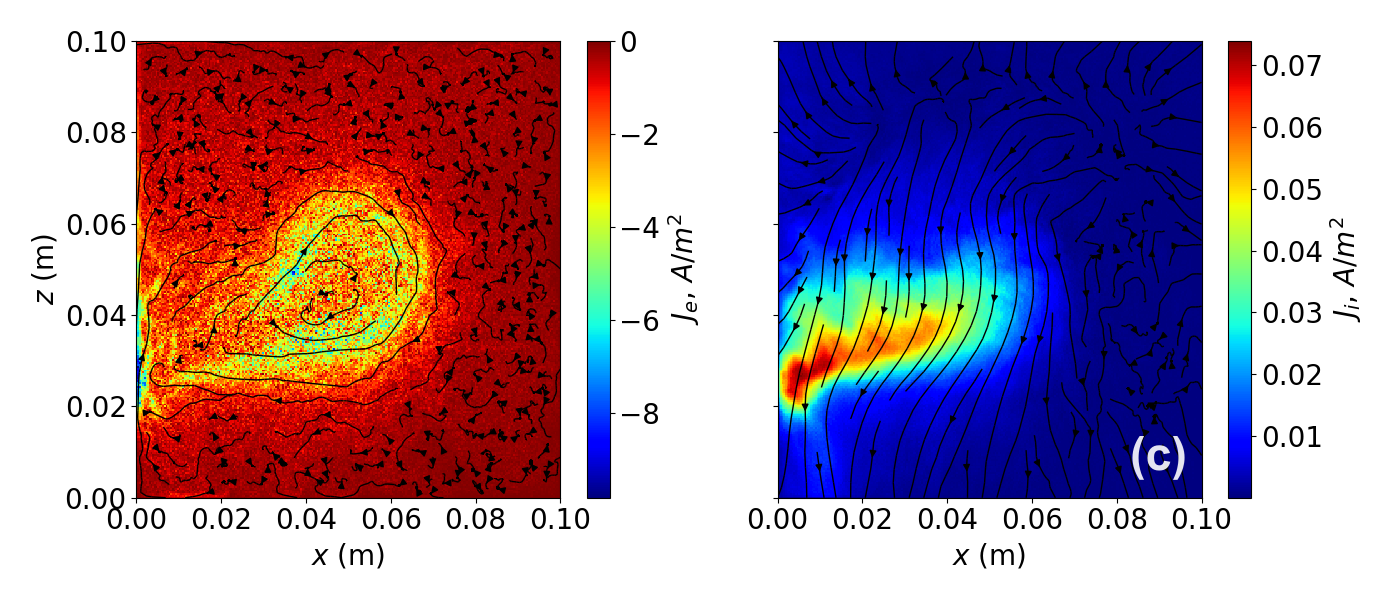}}\hfill
\vspace{-1.4cm}  
\subcaptionbox{\label{fig:current-220g}}{\includegraphics[width=0.8\linewidth]{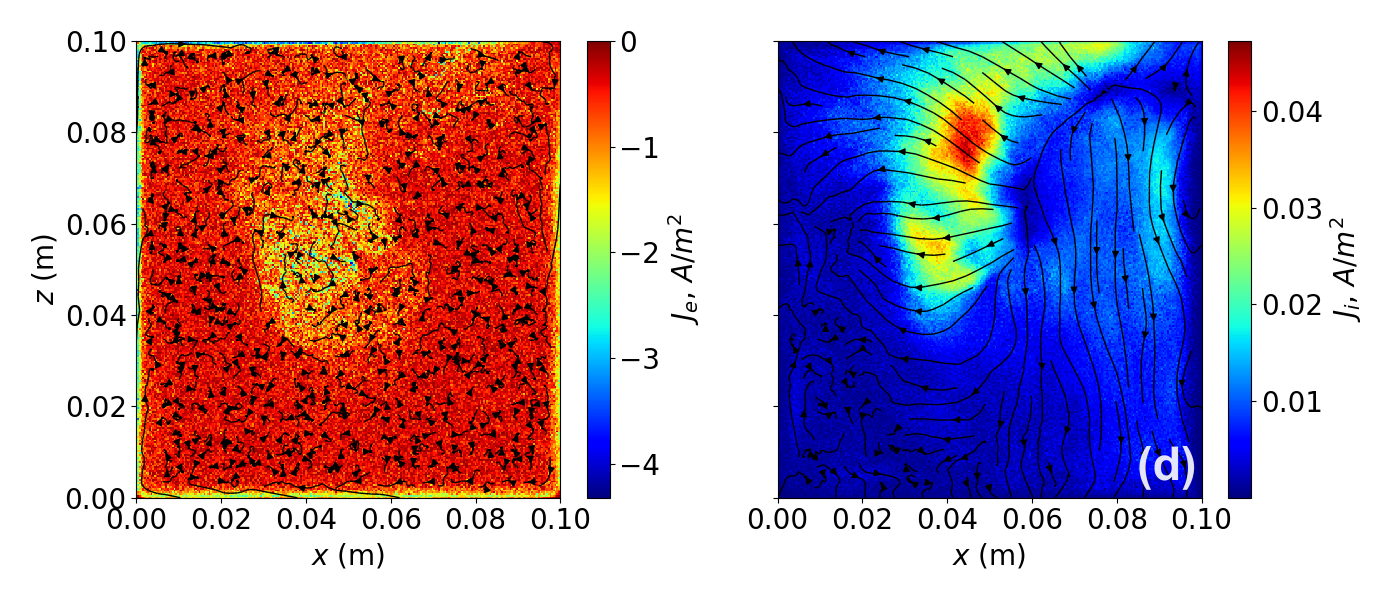}}\hfill
\vspace{-1.4cm}
\captionsetup{justification=raggedright,singlelinecheck=false}
\caption{Snapshots of the electron and ion current density, with arrows for the direction and colors for the current magnitude:  (a) $B=10\;$ G; (b) $B=50\;$ G; (c) $B=100\;$ G; (d) $B=220\;$ G; shown  at the same time moments as in Fig.\ref{fig:DensSnap}.}\label{fig:CurrentSnap}
\end{figure}

\begin{figure}[htp]
\centering
\captionsetup[subfigure]{labelformat=empty}
\subcaptionbox{\label{fig:}}{\includegraphics[width=0.8\linewidth]{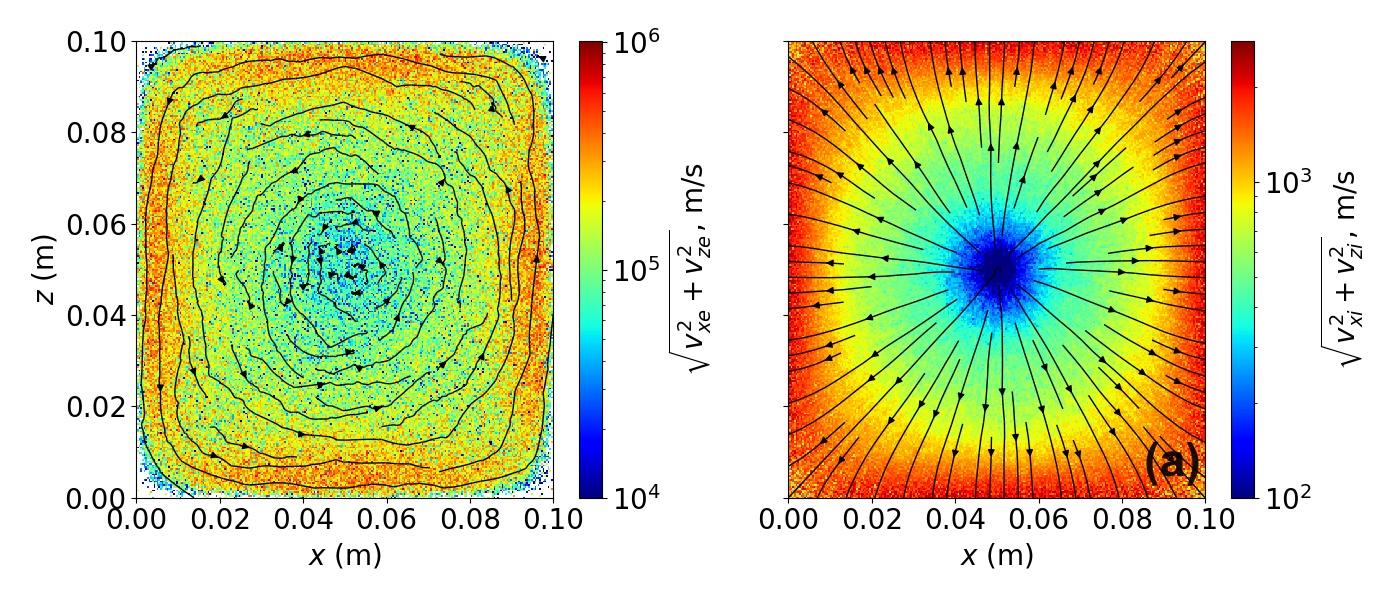}}\hfill
\vspace{-1.4cm}  
\subcaptionbox{\label{fig:}}{\includegraphics[width=0.8\linewidth]{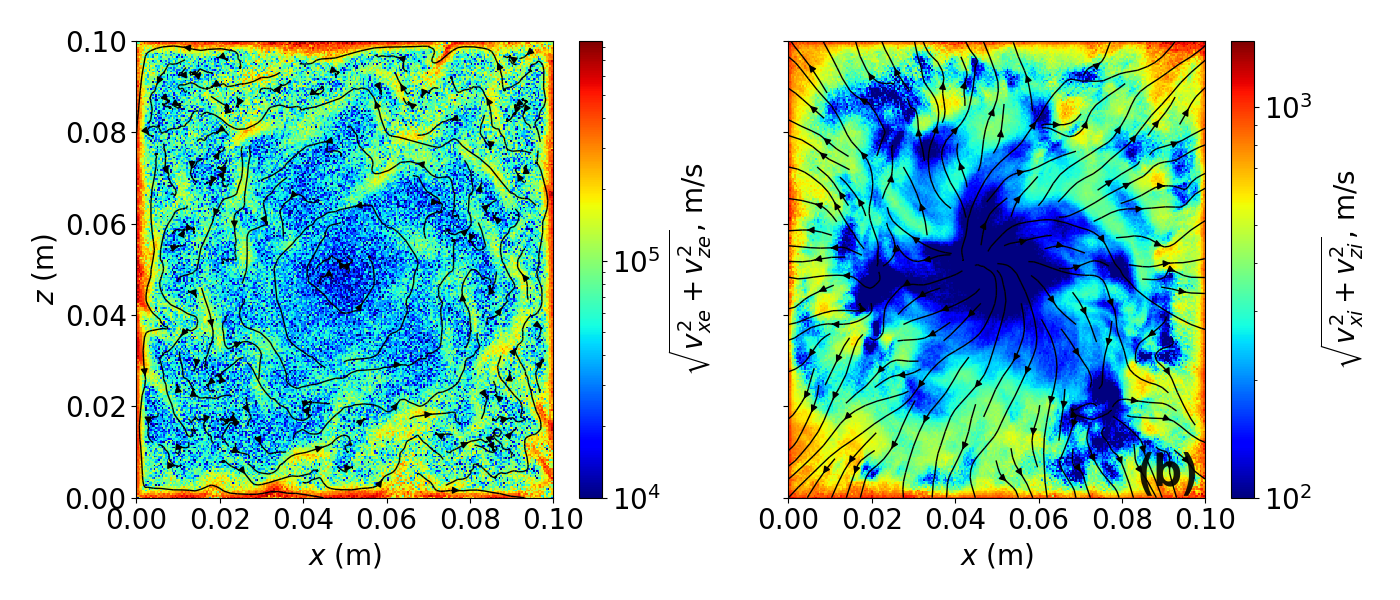}}\hfill
\vspace{-1.4cm}  
\subcaptionbox{\label{fig:}}{\includegraphics[width=0.8\linewidth]{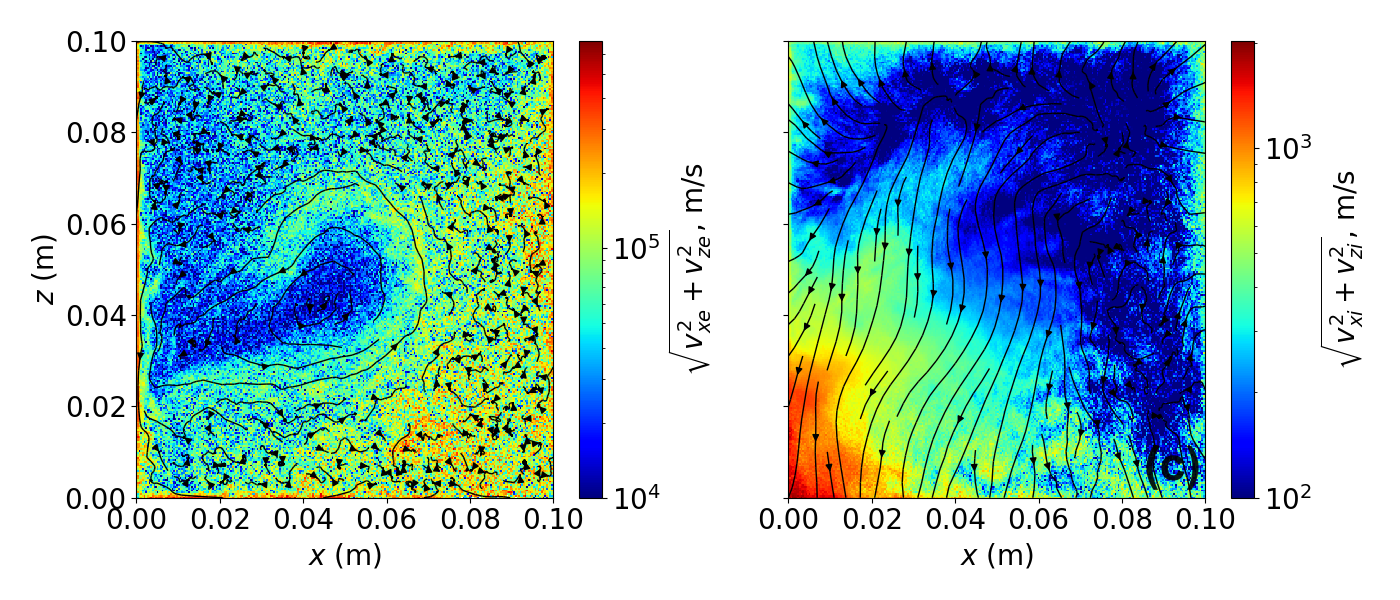}}\hfill
\vspace{-1.4cm}  
\subcaptionbox{\label{fig:}}{\includegraphics[width=0.8\linewidth]{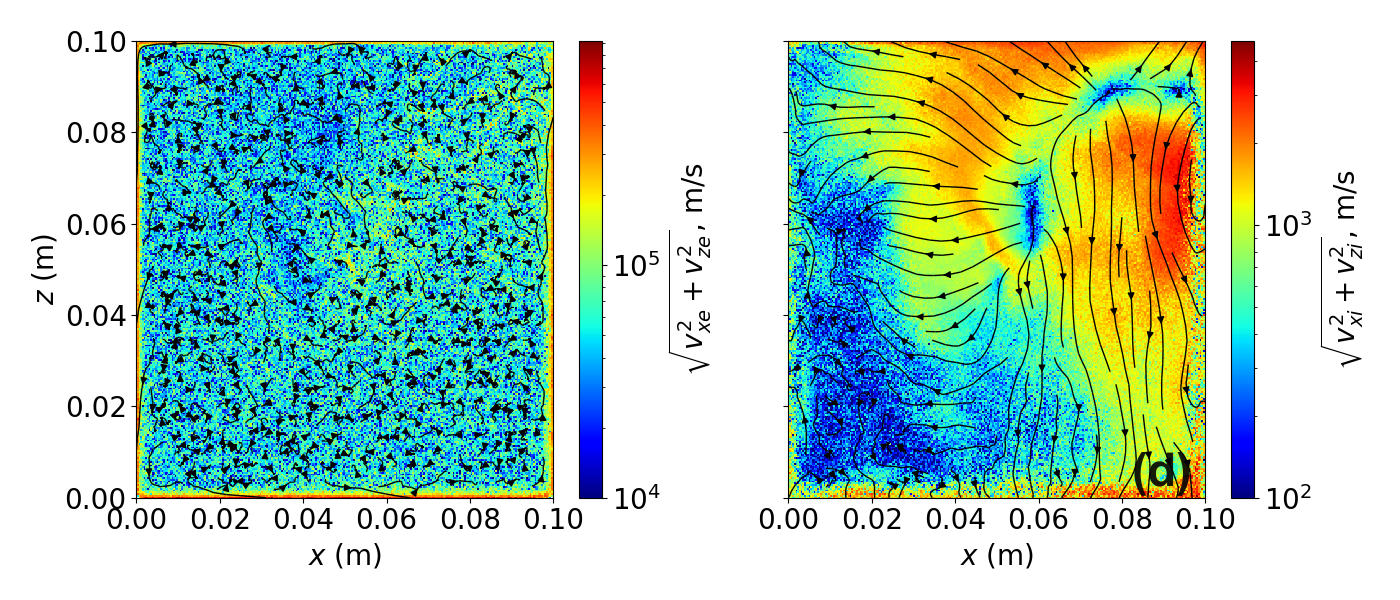}}\hfill
\vspace{-1.4cm}
\captionsetup{justification=raggedright,singlelinecheck=false}
\caption{Snapshots of the absolute value of the electron and ion velocity with the direction vectors for the different magnitude of the magnetic field: (a) $B=10\;$ G; (b) $B=50\;$ G; (c) $B=100\;$ G; (d) $B=220\;$ G; shown  at the same time moments as in Fig.\ref{fig:DensSnap}.}\label{fig:VelocitySnap}
\end{figure}

\begin{figure}[htp]
\centering
\captionsetup[subfigure]{labelformat=empty}
\subcaptionbox{\label{fig:}}{\includegraphics[width=0.8\linewidth]{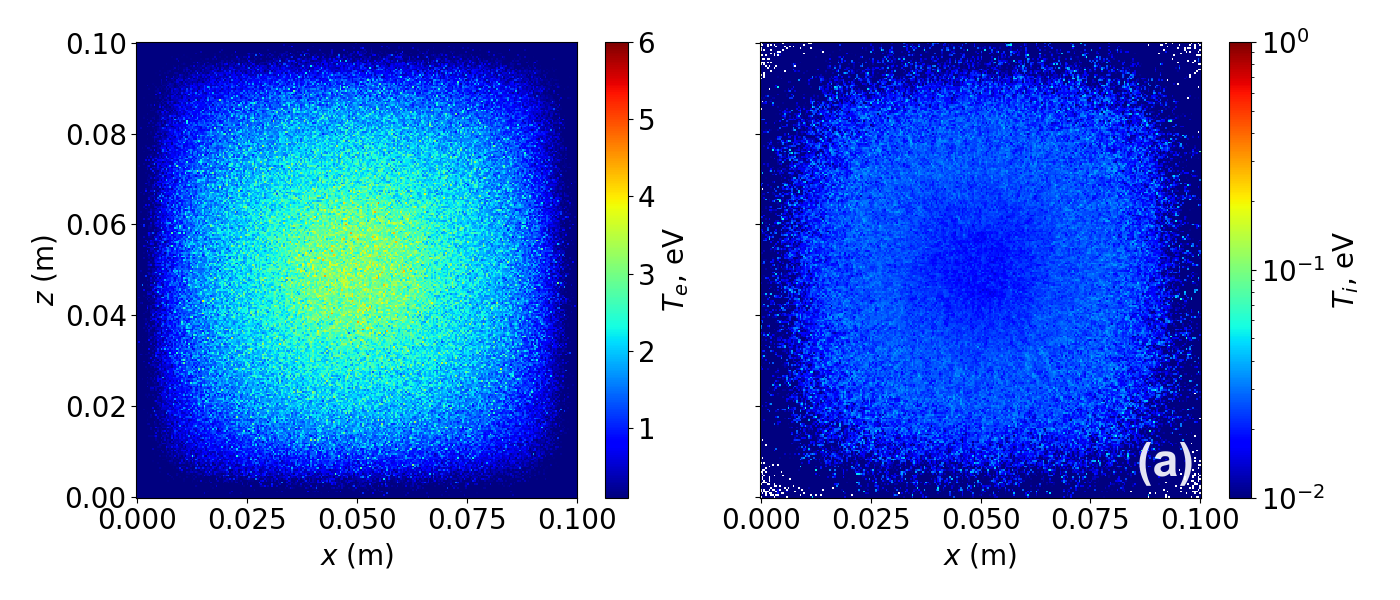}}\hfill
\vspace{-1.4cm}
\subcaptionbox{\label{fig:6c}}{\includegraphics[width=0.8\linewidth]{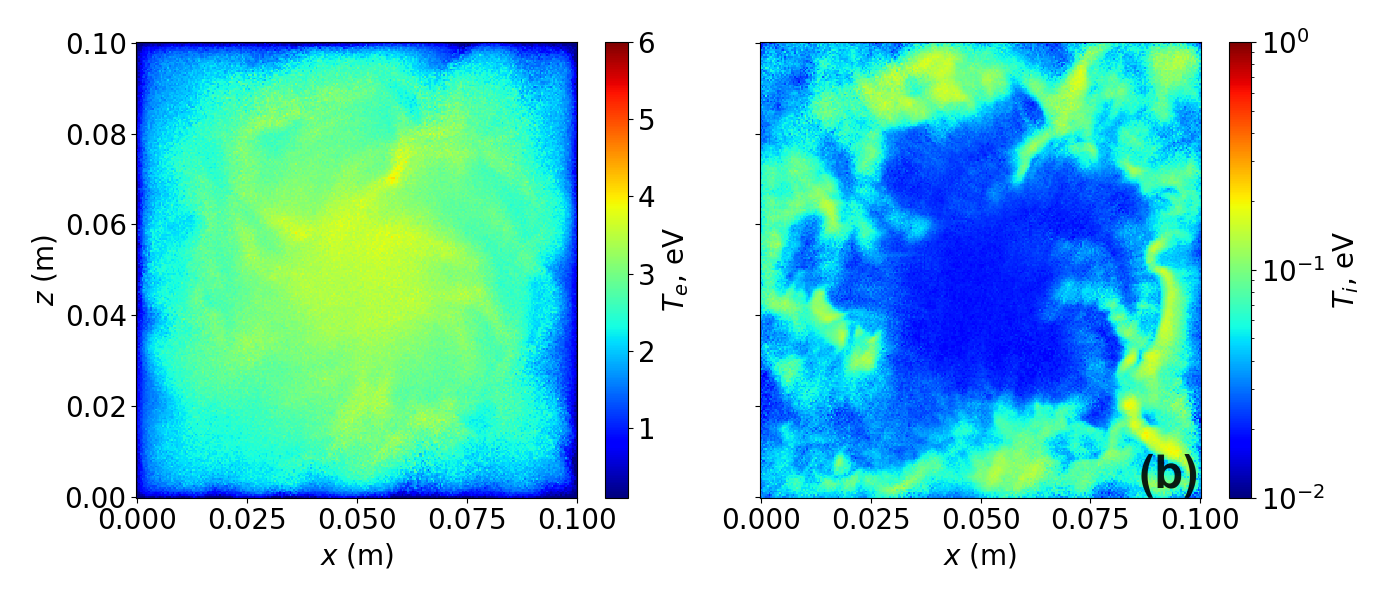}}\hfill
\vspace{-1.4cm}
\subcaptionbox{\label{fig:}}{\includegraphics[width=0.8\linewidth]{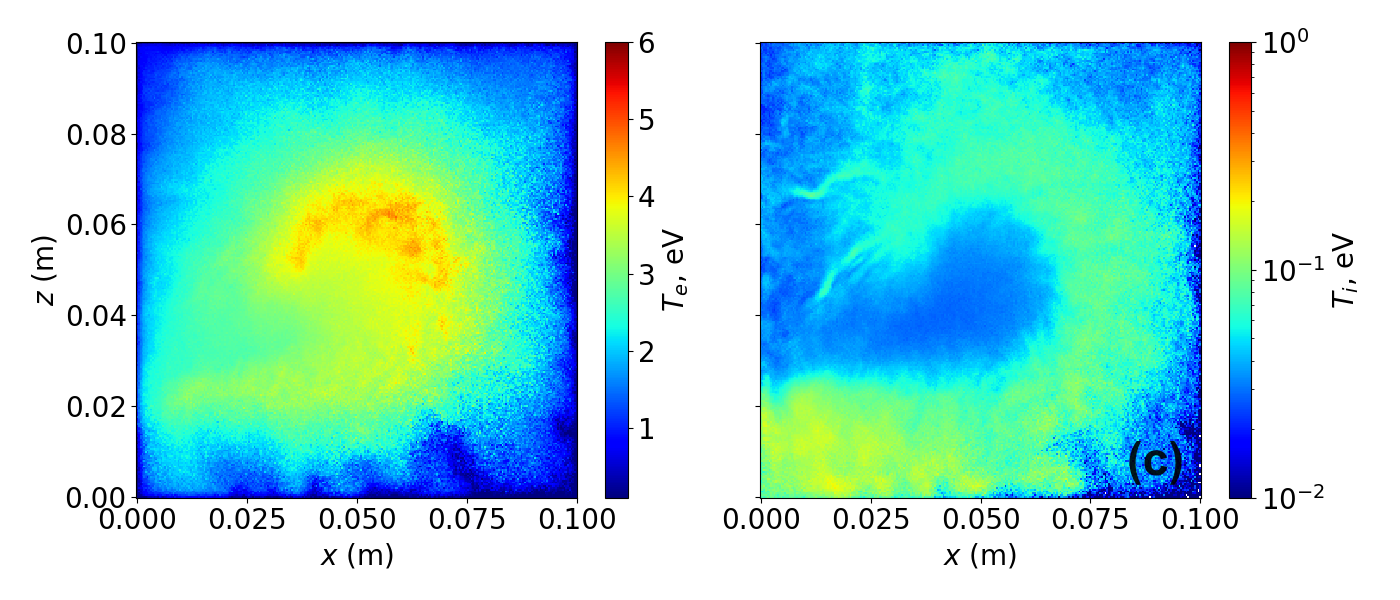}}\hfill
\vspace{-1.4cm}
\subcaptionbox{\label{fig:6d}}{\includegraphics[width=0.8\linewidth]{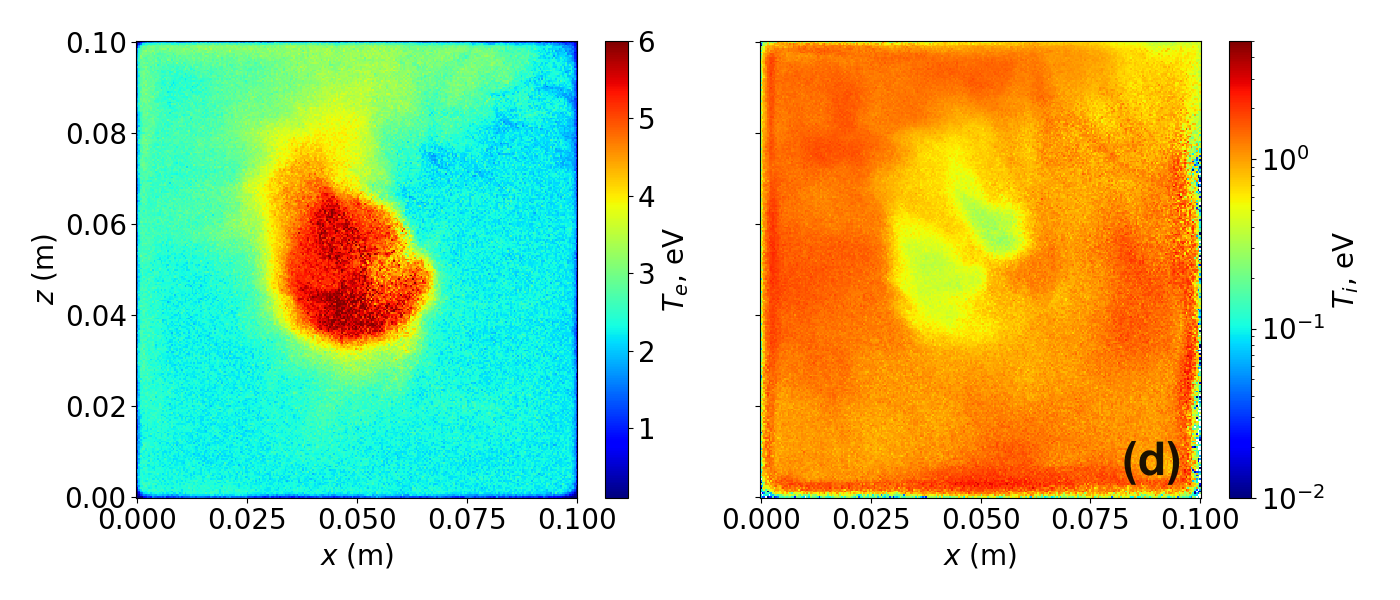}}\hfill
\vspace{-1.0cm}
\captionsetup{justification=raggedright,singlelinecheck=false}
\caption{Snapshots of the electron (left) and ion (ion) temperatures for the different magnitude of the magnetic field: (a) $B=10\;$ G; (b) $B=50\;$ G; (c) $B=100\;$ G; (d) $B=220\;$ G; shown  at the same time moments as in Fig.\ref{fig:DensSnap}.}\label{fig:TempSnap}
\end{figure}

The transition from the quiescent state at $B=10\;$G to the spiral arms and then to the spoke at a higher magnetic field is a result of marked changes in the radial electric field.  For low magnetic field values ($B=10$, and $B=50$ G), the radial electric field is zero or mildly positive while for the stronger field the radial electric field changes sign and becomes negative (inward). Figures \ref{fig:DensSnap} and \ref{fig:CurrentSnap}- \ref{fig:TempSnap} present the snapshots of the characteristic behavior of plasma parameters in the transition from spiral arm structure to spoke structure.

The spiral arms structures appear in the regime with zero or weakly positive electric field as it was also shown in our earlier simulations \cite{TyushevPoP2023}.
It was suggested in Ref. \onlinecite{LuckenPoP2019} that the spiral arms structures are the result of the resistive (dissipative) gradient drift instabilities. Such instabilities are essentially driven by the current perpendicular to the magnetic field and can be induced either by the electron or ion flow (or both). They belong to the class of negative energy perturbations that can become unstable due to the dissipative \cite{LashmoreJPP2005,Koshkarov2018PoPLH,LuckenPoP2019} or collisionless\cite{KoshkarovPoP2018ax} mechanisms.    Here, we suggest that the $m>1$ spiral arms structures 
in the low $B$ regime, Fig. \ref{fig:current-50g}, occur as a result of the ion radial motion. In the low $B$ case, in the stationary state,  the ions move radially outward due to a weak radial (outward) ambipolar electric field formed by the electron losses to the external boundary. The ion radial motion results in the spiral arms' perturbations moving radially and azimuthally in the direction of the electron diamagnetic drift. Spiral arms arise due to perturbations of density, which come from the combined radial motion of ions and density/potential perturbations propagating azimuthal. The azimuthal motion of ions occurs due to the azimuthal perturbation of potential, respectively, due to the azimuthal electric field.

% The generic condition for the instability is $v_c>c_s$\cite{KoshkarovPoP2018ax,LuckenPoP2019}, where   $v_c$ is the current velocity (in this case due to the ion motion) and $c_s=\sqrt{T_e/m_i}$. For the conditions in  Fig. \ref{fig:current-50g}, we have 
% This mechanism was discussed in Ref.   This instability results in fluctuations that propagate  in the direction of the electron diamagnetic drift and move radially as observed for the spiral arm structures in our simulations. The ion current streamlines in Fig. \ref{fig:Current-10g} show the outward radial ion flow. In Fig. \ref{fig:current-50g}  the outward ion radial flow is distorted by the radial-azimuthal instability. 

An increase in the axial magnetic field enhances electron confinement, forming a potential well and resulting in the inward electric field that triggers the collisionless Simon-Hoh mode for $\mathbf{E} \cdot \nabla n_0 >0$ leading to the spoke excitation \cite{SakawaPFB1993,SakawaPRL1992, SmolyakovPPCF2017}, also consistent with our results in Ref. \onlinecite{TyushevPoP2023}.  The flow of ions, which is strongly controlled by the electric field,  remains coherent within the spoke structures as seen in Figs. \ref{fig:CurrentSnap} and \ref{fig:VelocitySnap}.
Unlike ions, strongly magnetized electrons are subject to $\mathbf{E} \cross \mathbf{B}$ and diamagnetic drifts. Their motion is less coherent, Figs. \ref{fig:CurrentSnap} and \ref{fig:VelocitySnap}.  
It is important to note that ions are weakly magnetized but still subject to the magnetic field’s Lorentz force, which results in the deflection of their trajectories from the direction of the electric field. The measure of the magnetization is characterized by the parameter $E/rB\omega_{ci,ce}$ \cite{GueroultPOP2017,AggarwalPoP2023}. Magnetization is strong when this parameter is small, as for the electrons, when $B=100$ G, $E=0.5$ V/cm, $r=3$ cm, $E/rB\omega_{ce}\simeq 10^{-4}$ they experience simple  $\mathbf{E}\times \mathbf{B}$ and diamagnetic drifts. For argon ions, $E/rB\omega_{ce}\simeq 7$ and ion motion is strongly affected by the inertia (centrifugal forces) and deviates from the  $\mathbf{E}\times \mathbf{B}$ velocity \cite{TyushevPoP2023}. For larger values of the magnetic field, ion magnetization may increase, however, the electric field, which is controlled by plasma transport,  is also increasing. As a result, in our simulations, for a larger magnetic field of  $B=220$ G, ions remain weakly magnetized. Effects of finite ion magnetization on the Simon-Hoh instability were considered in Refs.  \onlinecite{GueroultPOP2017,AggarwalPoP2023}.  

The electron and ion temperature in Fig. \ref{fig:TempSnap} follow the behavior of the observed structures. The improvement of electron confinement with the magnetic field results in a higher density of energetic electrons in the central part of the discharge, Figs. \ref{fig:6c} and \ref{fig:6d}. The ions are collisionless here and their effective temperature corresponds to the energy of trapped ions bouncing in the potential well.  The ion Larmor radius corresponding to the well depth is about $0.02\;$m or larger, which is in the order of the system radius of $0.05\;$m. Also, the potential well is not stationary: it is deformed and is rotating with the spoke motion so the ions escape radially as a result of the azimuthal (spoke) rotation driven by the azimuthal electric field.
The electron Larmor radius remains smaller than the system size, providing  effective magnetic confinement for electrons throughout the domain.

\section{Role of energy conservation in the spoke formation}\label{sec:RoleOfEnergyCons}

\begin{figure}[htp]
 \centering
 \captionsetup[subfigure]{}
 \subcaptionbox{\label{fig:}}{\includegraphics[width=1\linewidth]{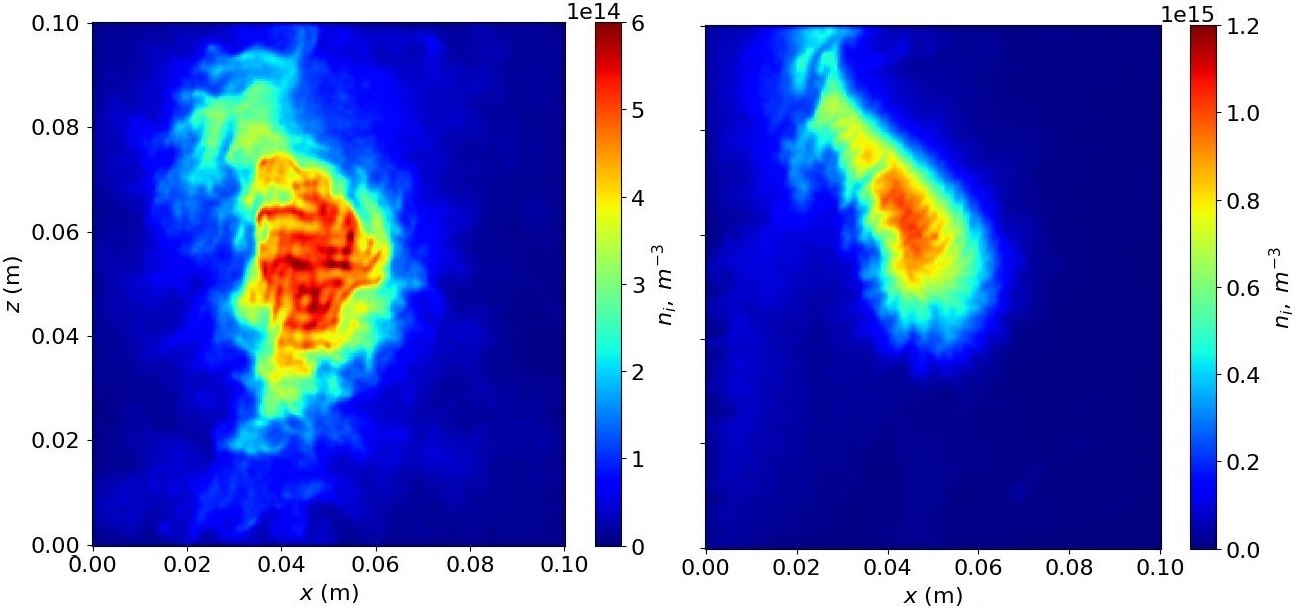}}
\vspace{-30pt}
\caption{The ion density perturbations in simulations using the collocated grid (left) at $5.68\times 10^{-3}\;$s and Yee grid (right) at  $4.9\times 10^{-3}\;$ s for $B=150 \;$G and $1 \;$mA current} 
\label{fig:checkers}
\end{figure}

Most of the simulations reported in this paper have been performed with the energy-conserving algorithm using the so-called Yee grid available in WarpX. For the electrostatic case, the main feature of the Yee grid is that the electric field is stored in the cell centers, contrary to the collocated grid where the electric field is defined on the nodes. The latter option, resulting in the momentum-conserving scheme is also available in WarpX.  
We have performed spoke simulations with both options, the collocated grid (momentum-conserving) and the Yee grid (energy-conserving). In the simulations with the momentum-conserving scheme, we observe unnatural checkered patterns at small scales and higher frequencies, and the shape of the $m=1$ large scale structures is also distorted as shown in Fig. \ref{fig:checkers} (left). It was found that these simulations have had energy errors of the order of $30-50 \%$.  Similar checkered patterns, spoke distortions, and energy errors were also observed in our situations with EDIPIC-2D which uses the momentum-conserving scheme (these results are not reported here). The simulations with an energy-conserving Yee grid demonstrate a much clearer picture of $m=1$ structure and small-scale fluctuations inside.  It is interesting to note that the energy errors appeared in the case when the $m=1$ spoke and small-scale fluctuations coexist. We have not observed energy conservation problems in our earlier EDIPIC simulations\cite{TyushevPoP2023} of the spoke in high neutral pressure regimes which did not have appreciable levels of coexisting small-scale structures.  Similar spoke behavior and energy errors in simulations with the momentum-conserving schemes were found  in simulations with two independent PIC codes performed at LAPLACE laboratory \cite{GECFubiani}. It was also found in these simulations \cite{GECFubiani} that higher resolution in momentum-conserving schemes improves energy conservation but does not completely solve the problem. Accuracy and advantages of the energy-conserving PIC algorithms were discussed in recent works\cite{PowisPoP2024,BarnesCPC2021}.

\section{Coexisting spoke and high-frequency small-scale structures.}\label{sec:100mA}

\begin{figure}[htp]
 \centering
 \captionsetup[subfigure]{}
 \subcaptionbox{\label{fig:}}{\includegraphics[width=1\linewidth]{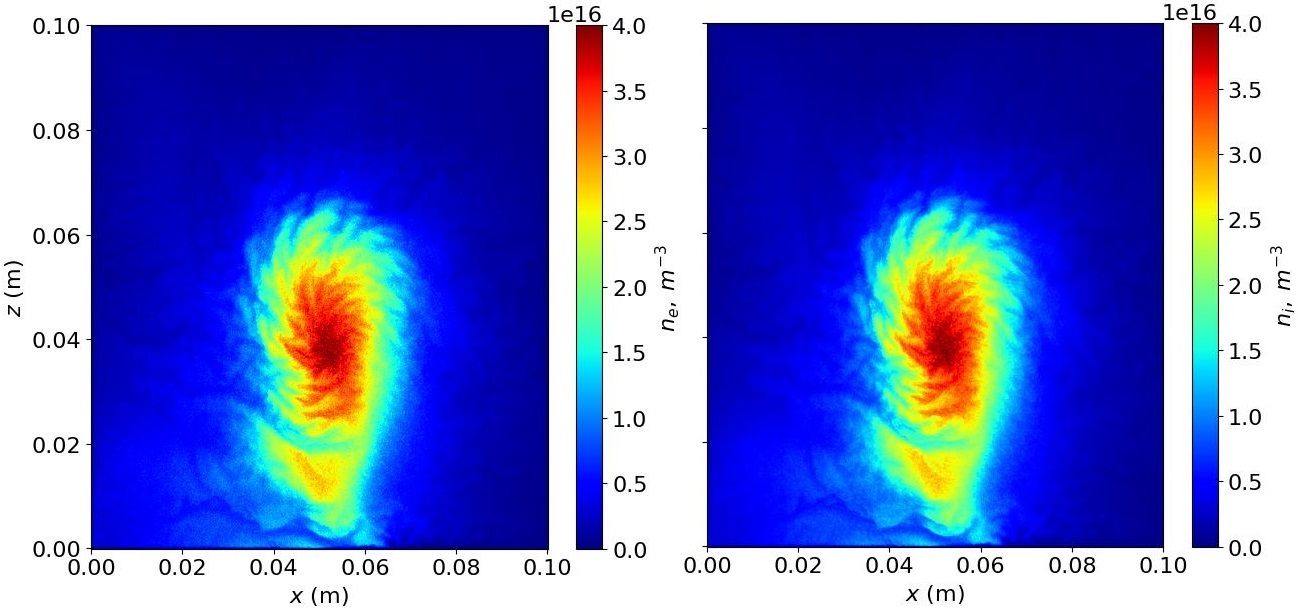}}
\vspace{-30pt}
\captionsetup{justification=raggedright,singlelinecheck=false}
\caption{High-frequency small-scale structures inside and on the periphery of the $m=1$ spoke in the high current simulations $100\;$ mA, $B=150\;$ G;  electron (left) and ion (right) density.} 
\label{fig:100_Spoke}
\end{figure}

\begin{figure}[htp]
\centering
\captionsetup{justification=raggedright,singlelinecheck=false}
\includegraphics[width=0.9\textwidth]{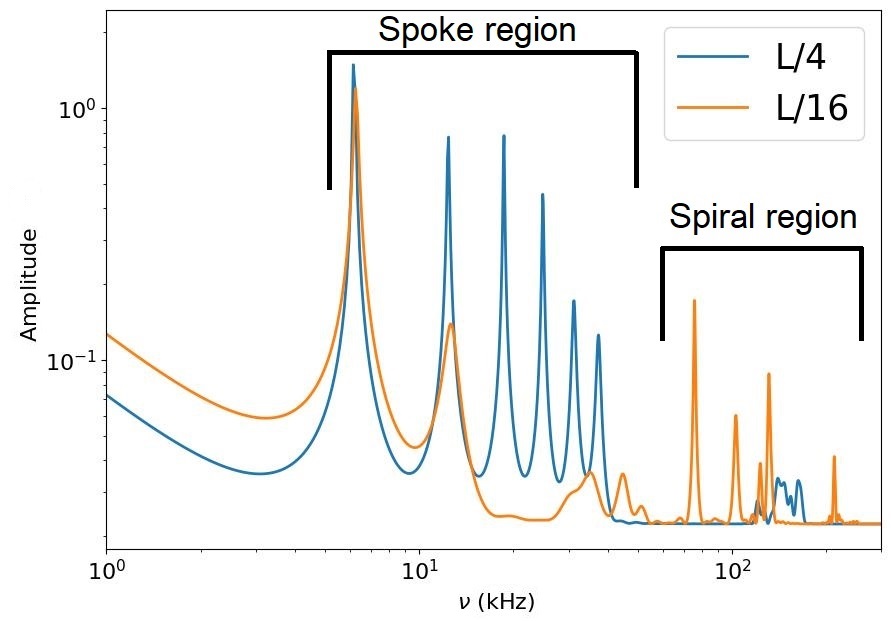}
\vspace{-10pt}
\caption{MUSIC spectra for probes at $L/4$ and $L/16$ distance from the center for simulations shown in Fig. \ref{fig:100_Spoke}.}
\label{fig:Music}
\end{figure}

\begin{figure}[htp]
 \centering
 \captionsetup[subfigure]{}
 \subcaptionbox{\label{fig:}}{\includegraphics[width=1\linewidth]{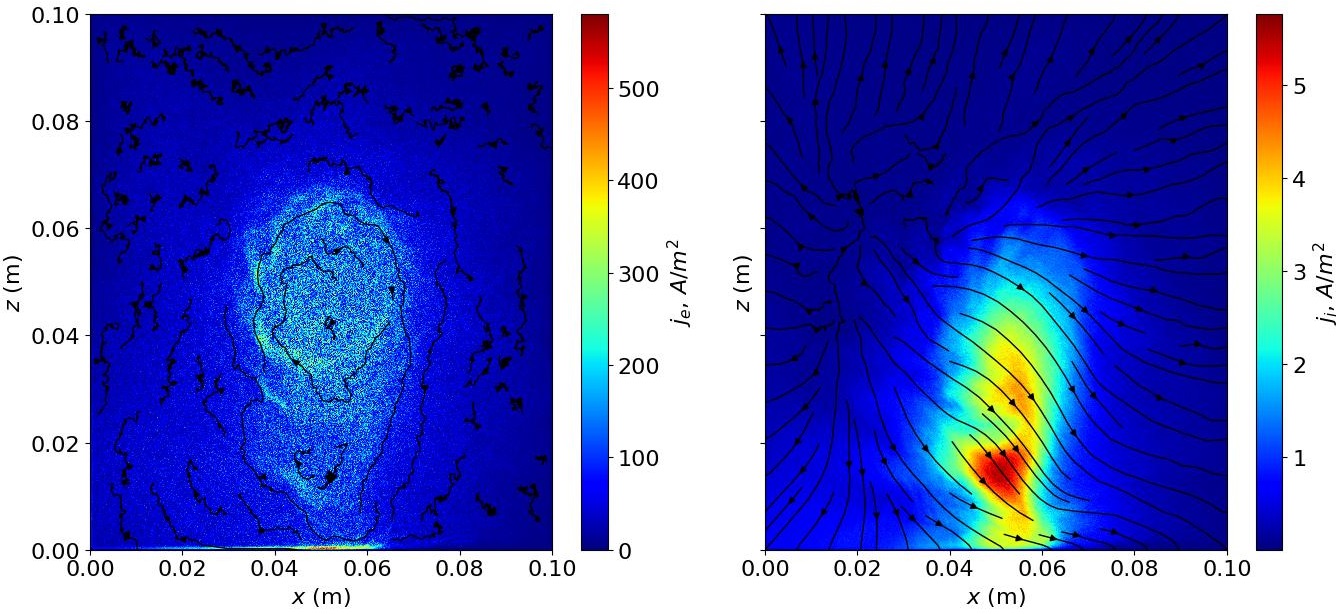}}
\vspace{-30pt}
\caption{Electron and ion current densities in the high current spoke simulations $100\;\text{mA},\; B=150\; \text{G}$.}
\label{fig:spoke100_currents}
\end{figure}

\begin{figure}[htp]
 \centering
 \captionsetup[subfigure]{}
 \subcaptionbox{\label{fig:}}{\includegraphics[width=1\linewidth]{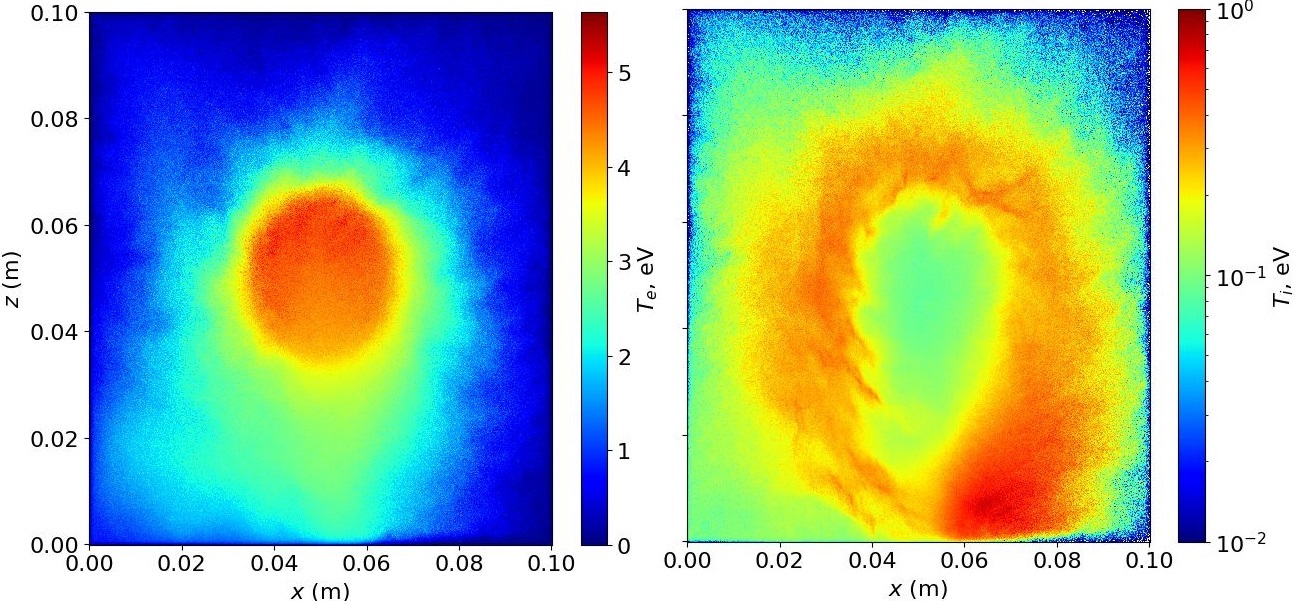}}
\vspace{-30pt}
\caption{Electron and ion temperatures in the high current spoke simulations $100\;\text{mA}, \; B=150\;\text{G}$.}
\label{fig:spoke100_temperatures}
\end{figure}

\begin{figure}[htp]
 \centering
 \captionsetup[subfigure]{skip=-20pt}
 \subcaptionbox{\label{fig:}}{\includegraphics[width=0.49\linewidth]{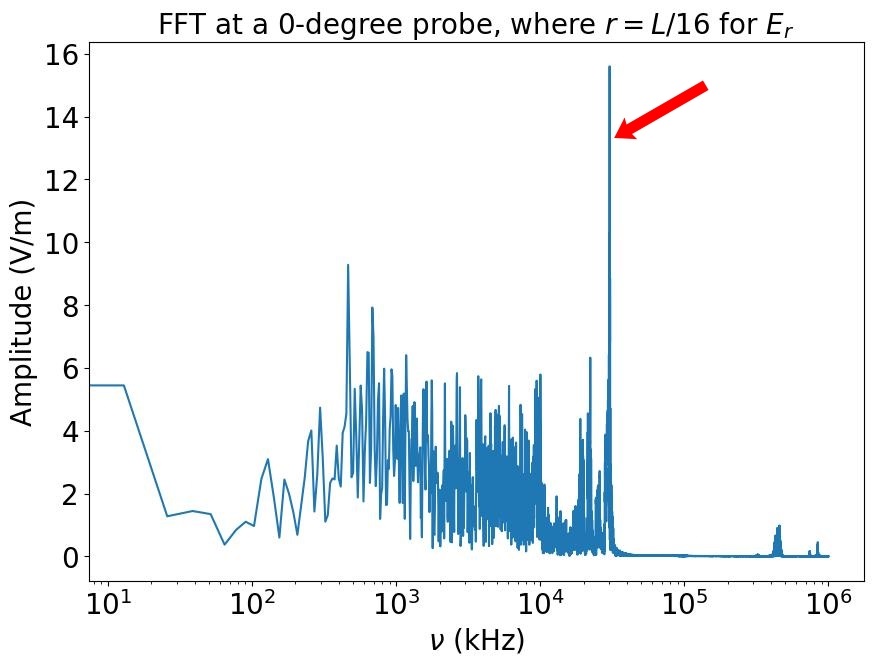}}
 \subcaptionbox{\label{}}{\includegraphics[width=0.49\linewidth]{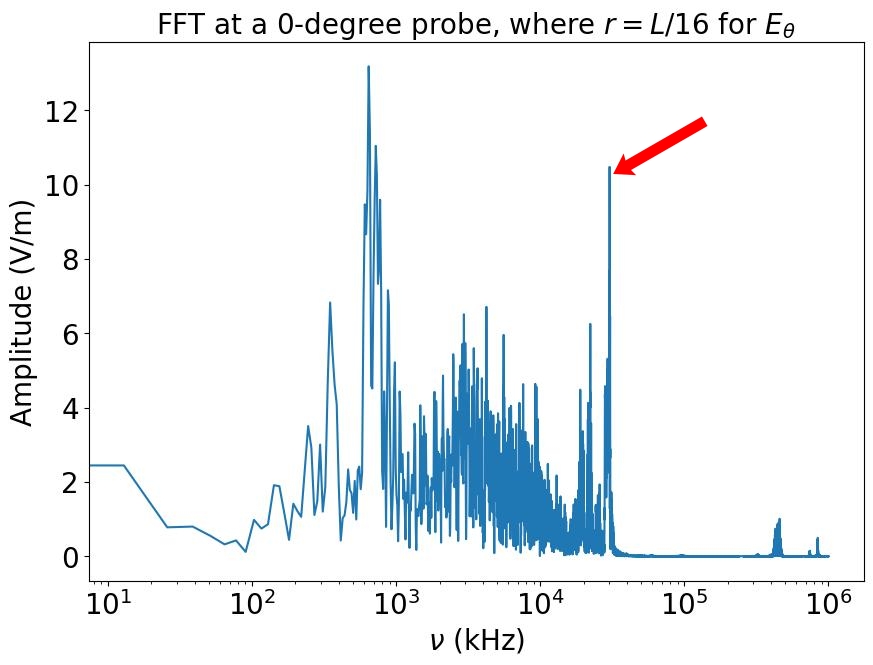}}
\captionsetup{justification=raggedright,singlelinecheck=false}
\vspace{-10pt}
\caption{Fourier spectra for $E_r$ and $E_\theta$ components of the electric field obtained from the probe data located at $\theta = 0$, $r = L/16$ from the center; the simulations with  $B=150 \; G$ and $1 \;$ mA.  The frequency at $3 \times 10^4$ kHz (shown by the red arrow) is consistent with the frequency of a lower-hybrid mode.}
\label{fig:LH}
\end{figure}

As it has been already noted above, the large scale $m=1$ spoke is accompanied by higher frequency small scale structures as is apparent in density snapshots in Fig. \ref{fig:DensSnap}c. The coexistence of spoke and high-frequency small-scale structures becomes even more pronounced in simulations with a larger value of the injected current of $100\;\text{mA}$ (high current case) as shown in Fig. \ref{fig:100_Spoke}. For this case, the simulation time and spatial steps were decreased to $dt = 10^{-11}$, and the mesh size is $1024 \times 1024$ to maintain the same CFL number. The weight of a single macroparticle for both species is 6250000 particles. The magnetic field strength is 150 G.
The small-scale fluctuations have a spiral arm structure and higher frequency. The MUSIC (multiple signal classification) \cite{hayes2009statistical} spectra analysis shows the frequencies of the density spiral arms structures in the range of $10^2$ kHz, Fig. \ref{fig:Music}. They are more pronounced in the region closer to the center of the discharge, at a distance of $L/16$ from the center. The density probe at a larger distance, $L/4$, shows intense harmonics of the main spoke frequency in the range of $10\;$ kHz and lesser amplitude of the $10^2$ kHz high-frequency modes. The general behavior of the electron and ion current and temperatures is shown in Figs. \ref{fig:spoke100_currents} and \ref{fig:spoke100_temperatures}, is similar to the base case with the current of $1\;$ mA (low current case).  One should note an elevated ion temperature in Fig. \ref{fig:spoke100_temperatures} on the outside of the spoke structure. We believe this is a result of the ion heating from the small-scale high-frequency lower-hybrid fluctuations which are destabilized by the density gradients across the spoke boundaries. The presence of fluctuations in the range of the lower-hybrid frequency is confirmed by Fast Fourier Transform spectra of the radial and azimuthal electric field obtained from the probe data as shown in Fig. \ref{fig:LH}.  

% \begin{figure}[H]
%  \centering
%  \captionsetup[subfigure]{}
%  \subcaptionbox{\label{fig:}}{\includegraphics[width=0.49\linewidth]{Mikhail_s_images/100_mA/pot_100mA.jpeg}}
%  \subcaptionbox{\label{}}{\includegraphics[width=0.47\linewidth]{Mikhail_s_images/100_mA/E_field_100mA.jpeg}}
% \captionsetup{justification=raggedright,singlelinecheck=false}
% \vspace{-30pt}
% \caption{Potential and electric field (7.83e-4 s).}
% \end{figure}


\section{Spoke structure in simulations with external plasma sources and the absence of e-n collisions}\label{sec:NoIoniz}

In many cases, the ionization processes are essential in the self-organization of plasma discharges resulting in the formation of nonlinear structures. This may occur either directly through the nonlinear dependencies of the ionization on electron temperatures (as in the formation of strata and striations \cite{BoeufPoP2022}) or via additional turbulence caused by instabilities resulting from the interactions between plasma and neutral flows. The latter mechanism is responsible for the phenomena of Critical Ionization Velocity (CIV), invoked for the spoke formation in some conditions \cite{BrenningSSR1992}.

We have tested earlier a possibility of spoke formation due to CIV in our earlier spoke simulations in high-pressure regimes \cite{TyushevPoP2023} where it was found that the spoke velocity is well below the CIV values.  
It is further suggested here that the ionization does not directly cause the mechanism of the spoke excitation. This has been confirmed here by the simulations in the absence of ionization where plasma discharge is supported by external sources of electrons and ions. 

 For these simulations,  we considered the regime with $1\;$ mA and $150\;$ G which has reached a steady state. At this stage, the electron and ion currents to the walls were $I_e = 2 \; \text{mA}$  and $I_i = 1 \; \text{mA}$ , respectively. The electron and ion temperature established in the injection region were $T_e = 5 \; \text{eV}$ and $T_i = 0.1 \; \text{eV}$. Then,  all electron-neutral collisions were turned off, and electrons and ions were generated in the circular region (as in Fig.\ref{fig:SimulationSetUp}) at the rates corresponding to their currents and with the temperatures equal to their stationary values as above. 

Figures  13-16 show the comparison of plasma parameters in the simulations with the ionization and with external sources in the absence of ionization. In general, one observes fairly similar behavior and structure patterns.

\begin{figure}[htp]
 \centering
 \captionsetup[subfigure]{skip=-20pt}
 
 \subcaptionbox{\label{fig:}}{\includegraphics[width=0.9\linewidth]{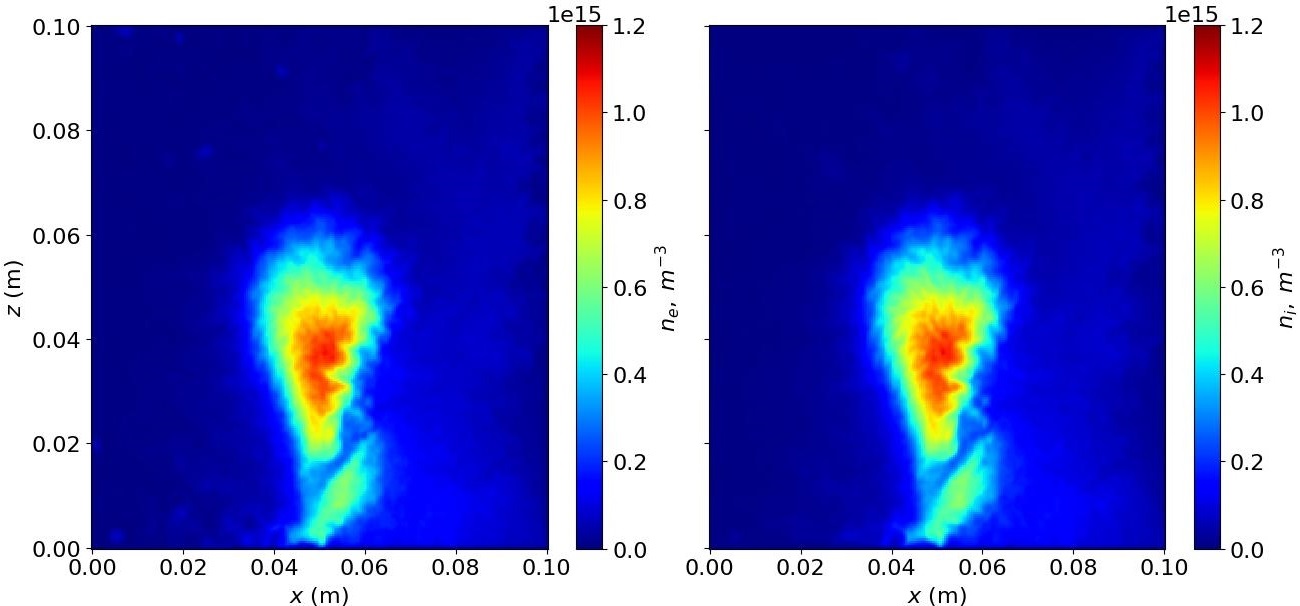}}

 \subcaptionbox{\label{}}{\includegraphics[width=0.9\linewidth]{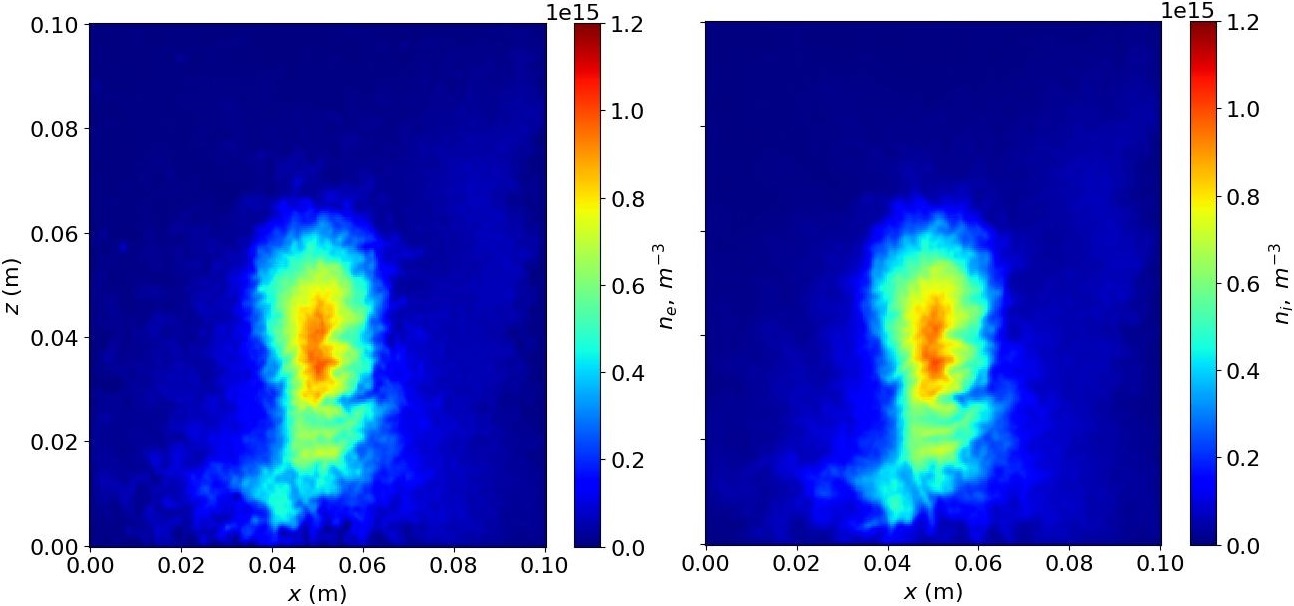}}
\captionsetup{justification=raggedright,singlelinecheck=false}
\vspace{-10pt}
\caption{Electron and ion concentrations for the regime with ionization (top row) at $5.05\times 10^{-3}\;$ s,  and with external sources (bottom row), at  $6.46\times 10^{-3}\;$ s for $B=150 \;$G and $1 \;$mA current.}
\end{figure}

\begin{figure}[htp]
 \centering
 \captionsetup[subfigure]{skip=-20pt}
 \subcaptionbox{\label{fig:}}{\includegraphics[width=0.49\linewidth]{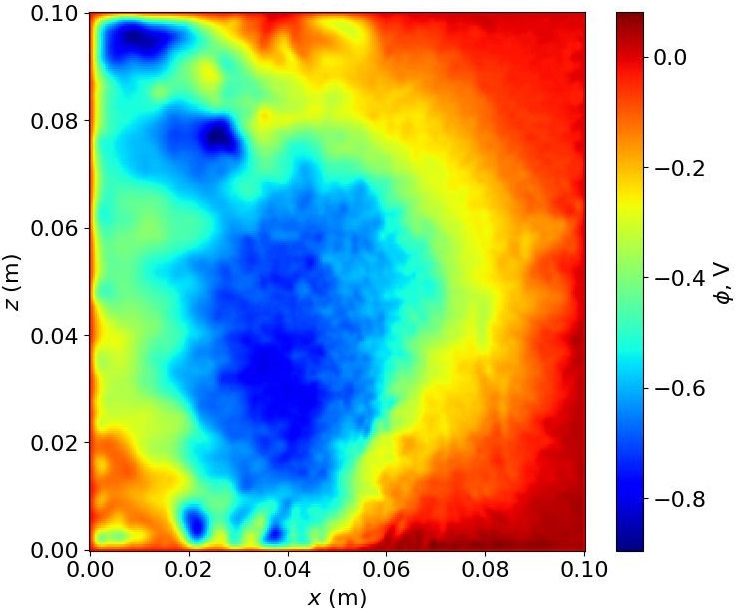}}
 \subcaptionbox{\label{}}{\includegraphics[width=0.43\linewidth]{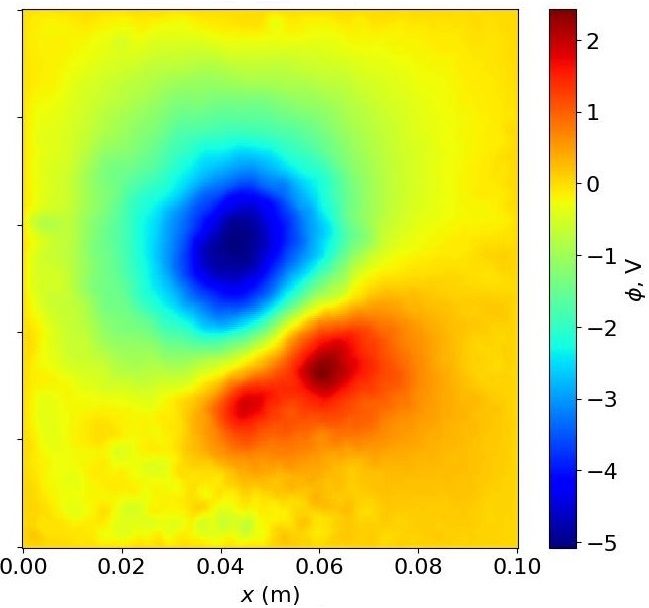}}
 {\includegraphics[width=0.48\linewidth]{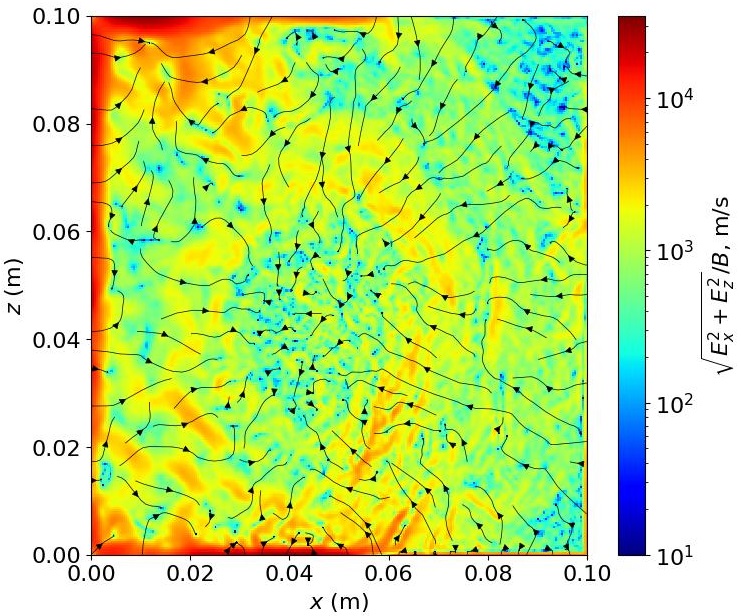}}
 \subcaptionbox{\label{}}{\includegraphics[width=0.48\linewidth]{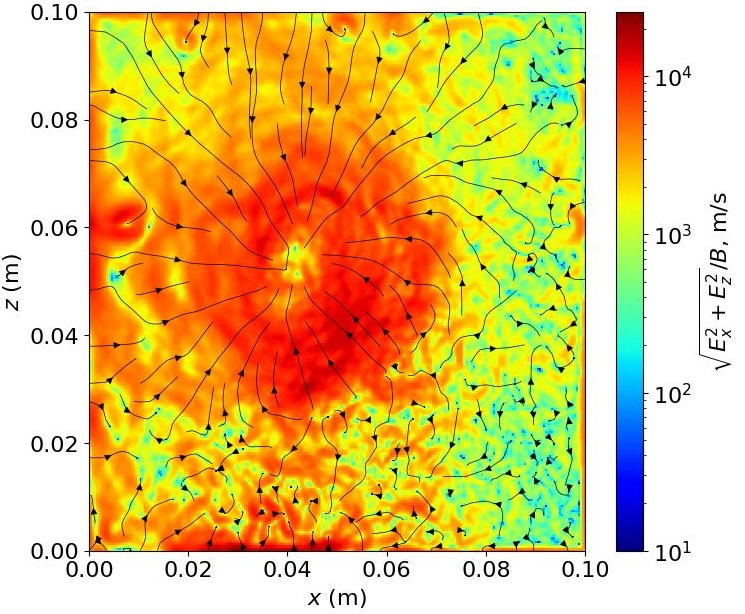}}
\captionsetup{justification=raggedright,singlelinecheck=false}
\vspace{-10pt}
\caption{Top row: Potential map for the regime with the ionization (left)  and with external sources  (right). Bottom row:  $\mathbf{E}\times \mathbf{B}$ velocity for the regime with the ionization and with external sources for $B=150 \;$ G and $1 \;$ mA current.}
\end{figure}

\begin{figure}[htp]
 \centering
 \captionsetup[subfigure]{skip=-20pt}
 \subcaptionbox{\label{fig:}}{\includegraphics[width=1\linewidth]{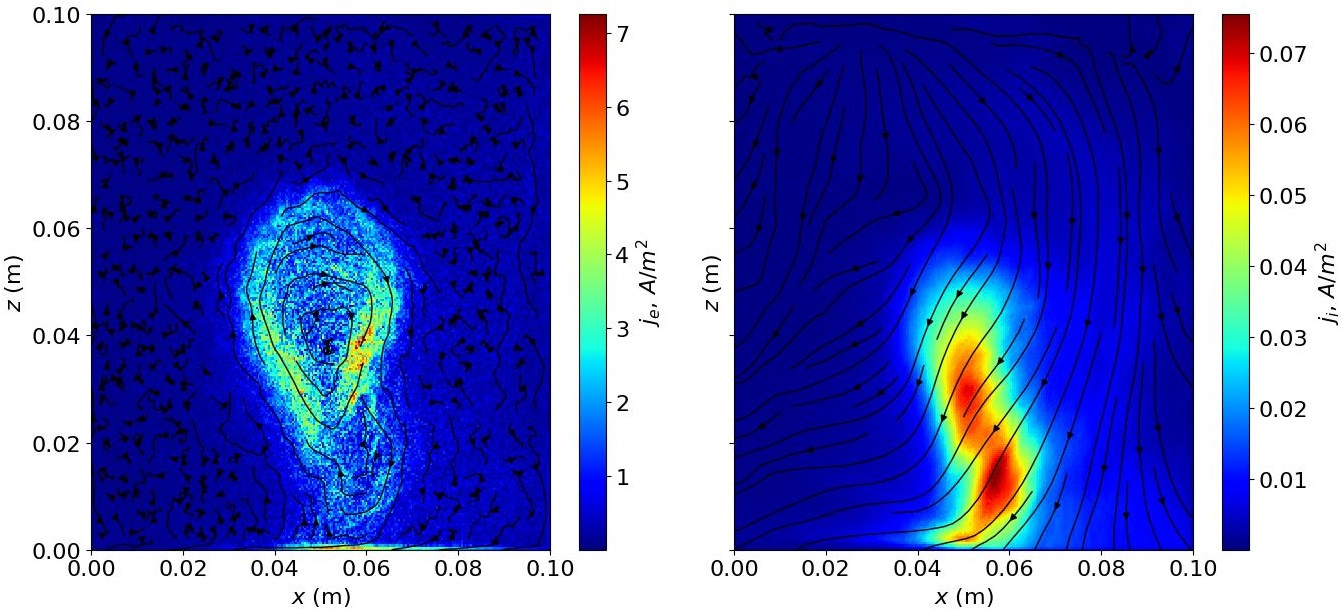}}
 \subcaptionbox{\label{}}{\includegraphics[width=1\linewidth]{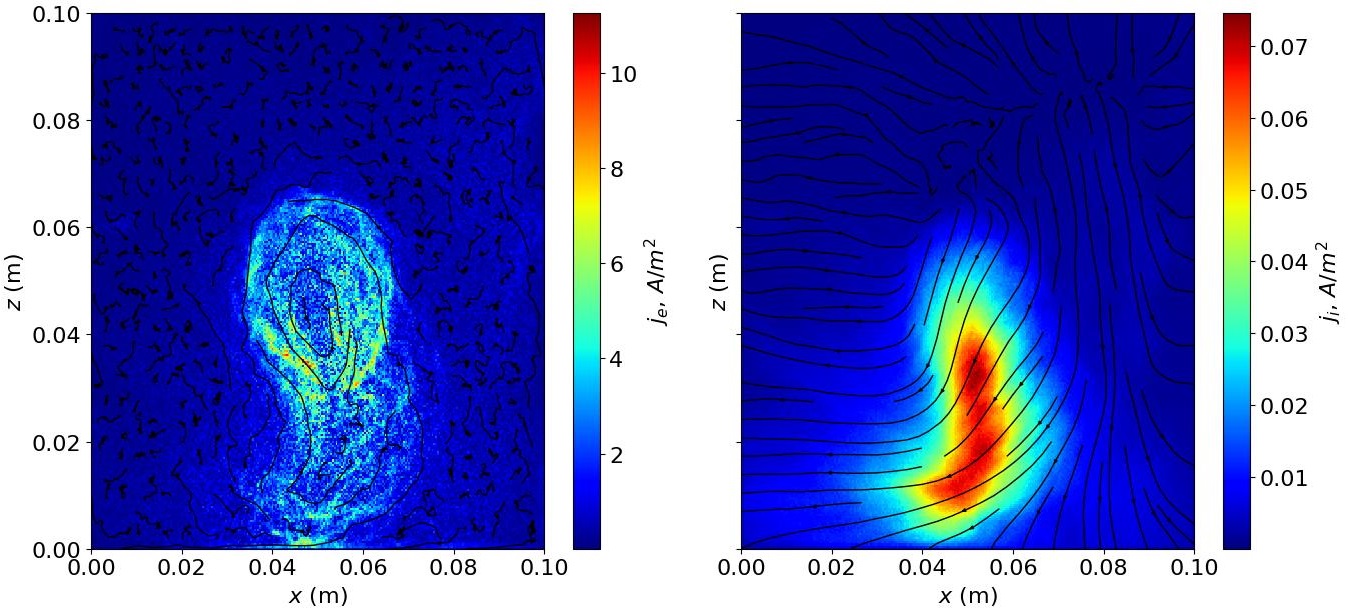}}
\captionsetup{justification=raggedright,singlelinecheck=false}
\vspace{-30pt}
\caption{ Electron and ion currents in the simulations with the ionization(top row) and with external sources (bottom row) for $B=150 \;$G and $1 \;$ mA current.}
\label{fig:streamlines}
\end{figure}

\begin{figure}[htp]
 \centering
 \captionsetup[subfigure]{skip=-20pt}
 \subcaptionbox{\label{fig:}}{\includegraphics[width=1\linewidth]{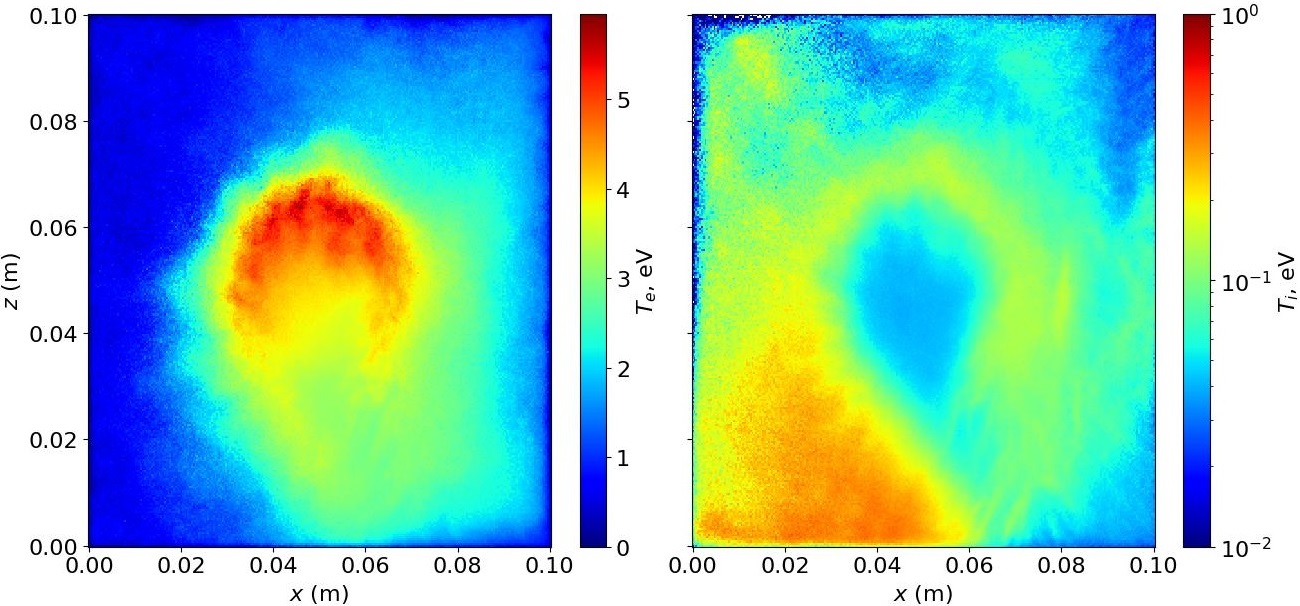}}
 \subcaptionbox{\label{}}{\includegraphics[width=1\linewidth]{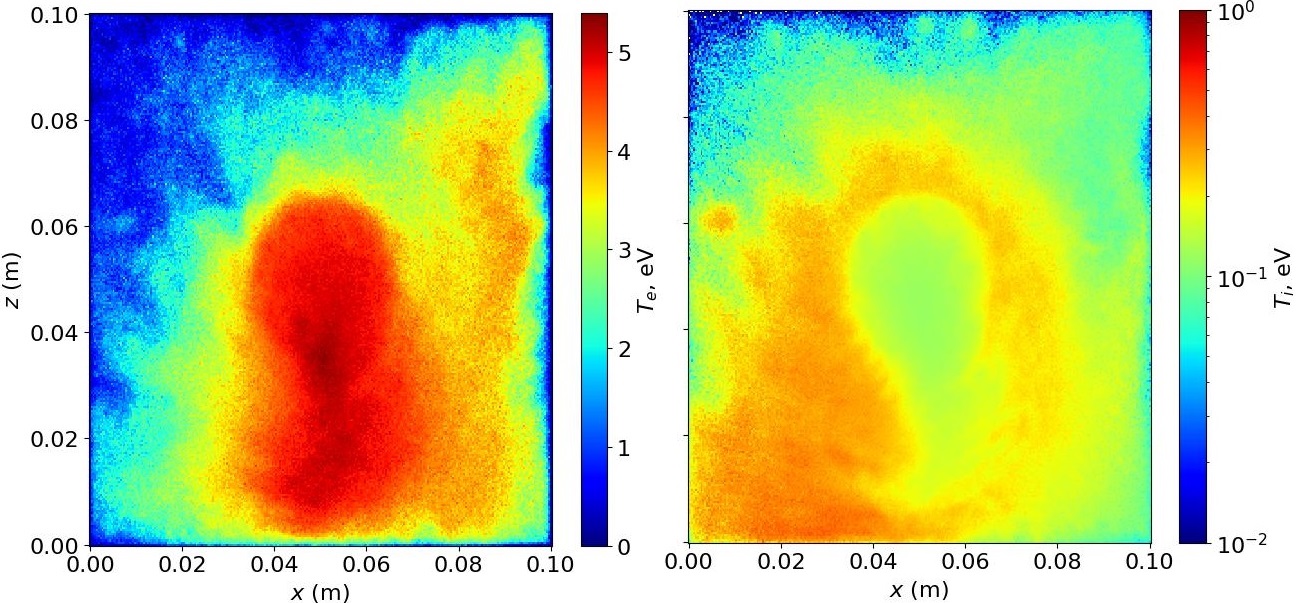}}
\captionsetup{justification=raggedright,singlelinecheck=false}
\vspace{-30pt}
\caption{Electron and ion temperatures in the simulations with the ionization (top row) and with external sources (bottom row) for $B=150 \;$G and $1 \;$ mA current.}
\end{figure}

\section{Effects of the geometry: oscillations in the ion and electron fluxes}\label{sec:EffectOfGeometry}

\begin{figure}[htp]
 \centering
 %\captionsetup[subfigure]{labelformat=empty}
 \captionsetup[subfigure]{}
 \subcaptionbox{\label{fig:CircleVsSquare-PartcileEv}}{\includegraphics[width=0.49\linewidth]{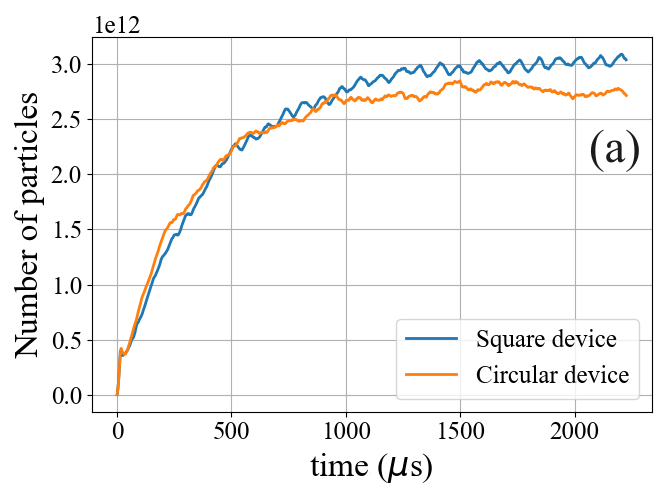}}
 \subcaptionbox{\label{fig:CircleVsSquare-fs}}{\includegraphics[width=0.49\linewidth]{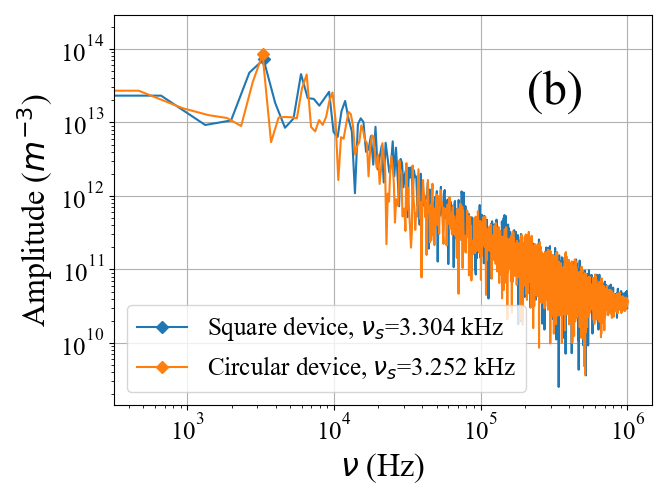}}
 \captionsetup{justification=raggedright,singlelinecheck=false}
 \vspace{-40pt}
\caption{(a) Temporal evolution of the total number of particles (electrons $+$ ions), and (b) spoke frequency of Argon gas in circular and square simulation geometry for $B=100 \;$ G and $1 \;$ mA current.
}\label{fig:CircleVsSquare-fsPartcileEvol}
\end{figure}

\begin{figure}[htp]
 \centering
 \captionsetup[subfigure]{skip=-30pt}
 \subcaptionbox{\label{fig:}}{\includegraphics[width=0.89\linewidth]{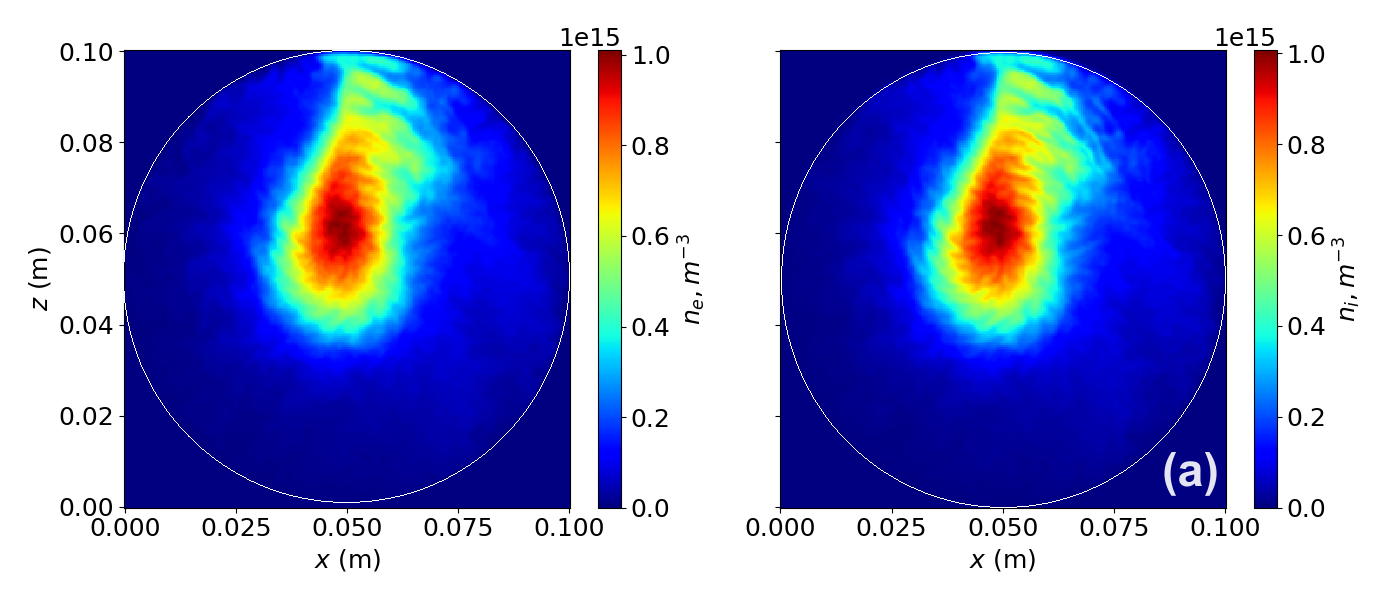}}
 \subcaptionbox{\label{}}{\includegraphics[width=0.89\linewidth]{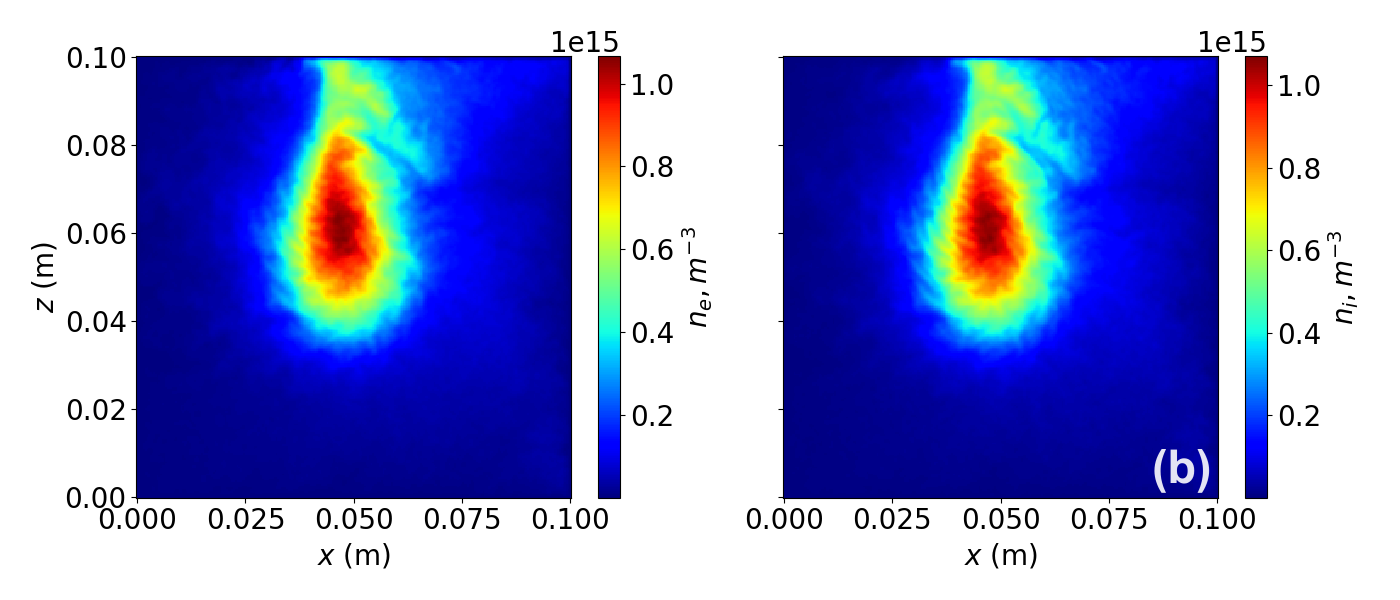}}
\captionsetup{justification=raggedright,singlelinecheck=false}
\vspace{-20pt}
\caption{The electron and ion densities for the $m=1$ spoke in  (a) circular, white line shows the location of the circular boundary and (b) square boundary geometry for $B=100 \;$ G and $1 \;$ mA injection current.}
\label{fig:CircleVsSquare}
\end{figure}

In our base case simulations with e-n collisions, we used a square box geometry of the external boundary.  Various applications of the discharges with axial magnetic fields may use both circular and rectangular shapes as well as their combinations for the external boundaries \cite{AbolmasovPSST2012, KimPSST2022, FubianiPoP2012,BoeufPoP2012b,JunePoP2023,JunePSST2023}.
Therefore,  it is of interest to explore the differences arising from the shape of the external boundaries. To change geometry to a circle we use what's called an embedded boundary object in WarpX. It allows the addition of curved boundaries to the system with rectangular geometry. In our case, we add a circle to our square geometry as an implicit function $(x - L_x/2)^2 + (z - L_x/2)^2 = (L_x/2)^2$ (see Fig. \ref{fig:SimulationSetUp}). This boundary has the same features as the rectangular boundary,  absorbing all incoming particles and zero potential as a boundary condition for the solution of the Poisson equation. Figure \ref{fig:CircleVsSquare-fsPartcileEvol} shows the evolution of the inventory of particles in the  Argon discharge simulation within a circular boundary, as in Fig. \ref{fig:SimulationSetUp}, and compares it with the case of the square boundary. One observes a small difference in the total number of particles at the saturation,  which can be explained by the difference in the plasma volume for the two geometries. 
The rotation frequencies of the spoke in square and circular geometries obtained from the Fourier Transform of the density perturbation in the $\theta-t$  are nearly identical as shown in Fig. \ref{fig:CircleVsSquare-fs}.  The general shape of the density structures in the $m=1$ spoke in the square and circular geometries are also very similar as shown in Figs. \ref{fig:CircleVsSquare}.

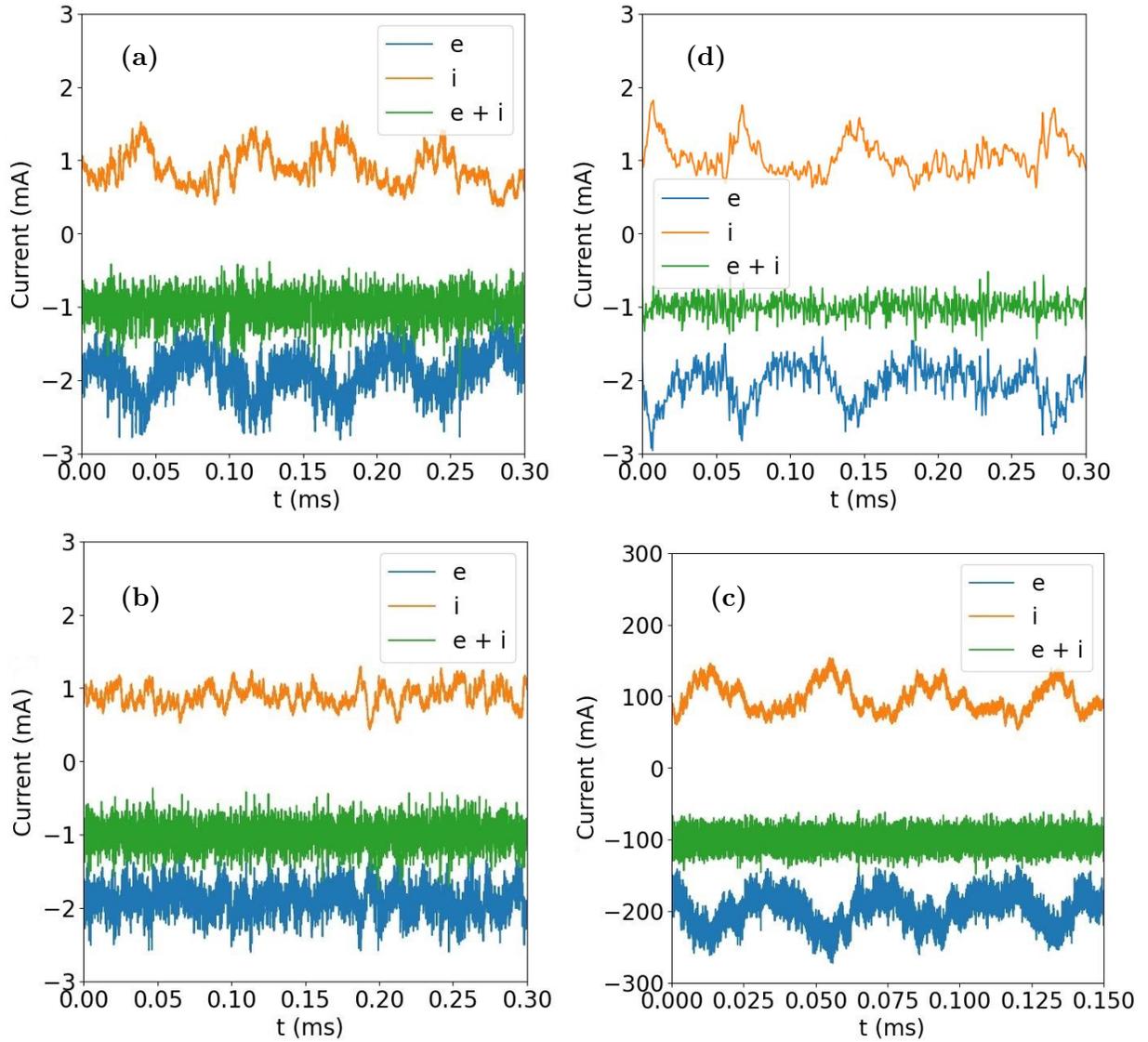
\begin{figure}[H]
\centering
\captionsetup[subfigure]{}
\subcaptionbox{\label{fig:}}

{\includegraphics[width=0.48\linewidth]{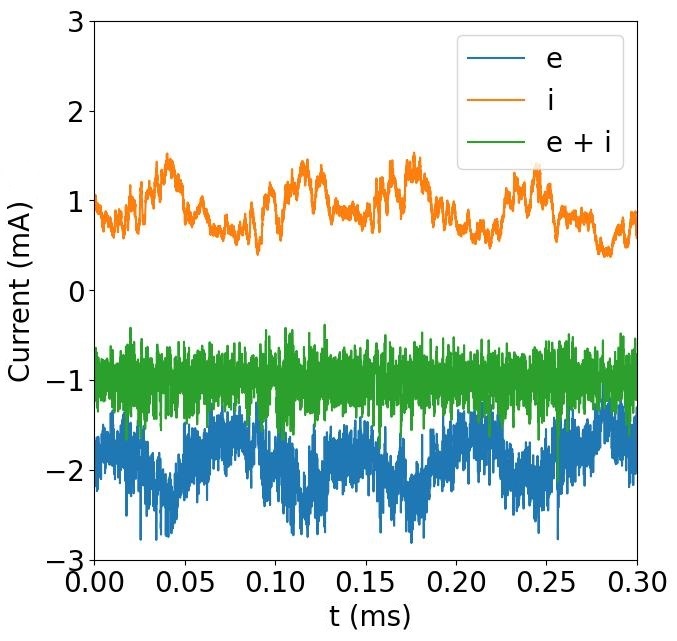}}
\subcaptionbox{\label{}}
{\includegraphics[width=0.48\linewidth]{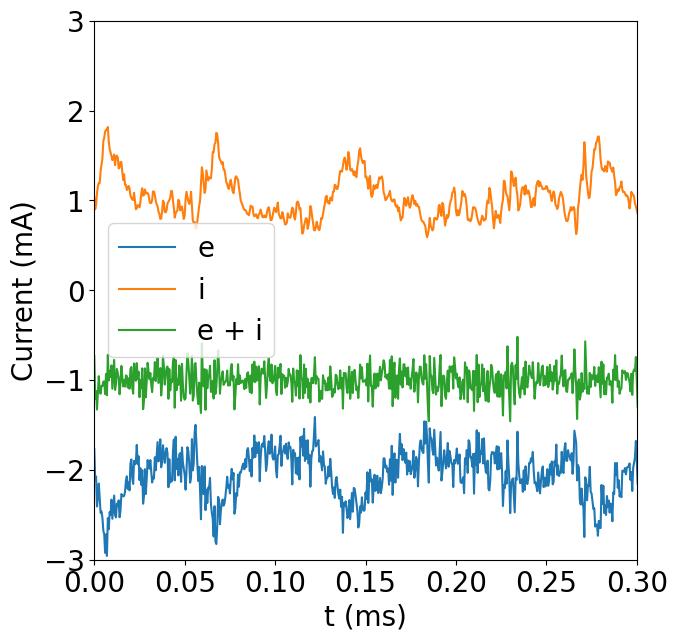}}
\hfill
\vspace{-1 cm}
\subcaptionbox{\label{}}
{\includegraphics[width=0.48\linewidth]{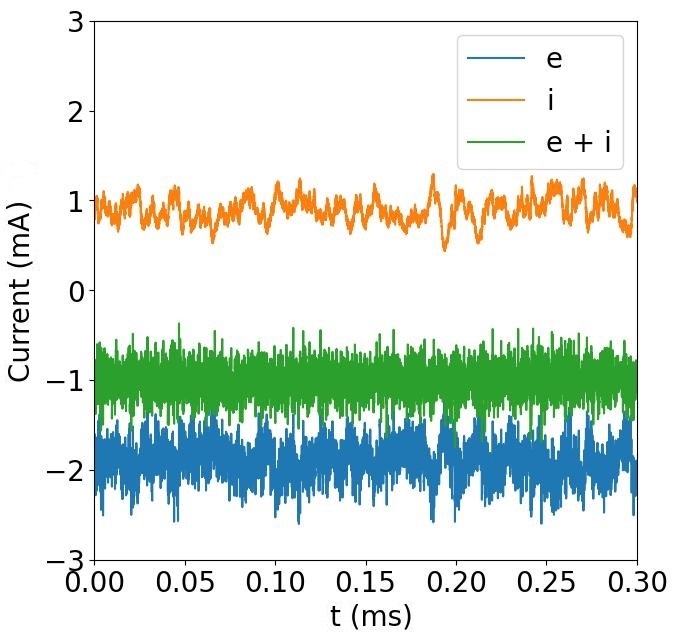}}
\subcaptionbox{\label{}}
{\includegraphics[width=0.5\linewidth]{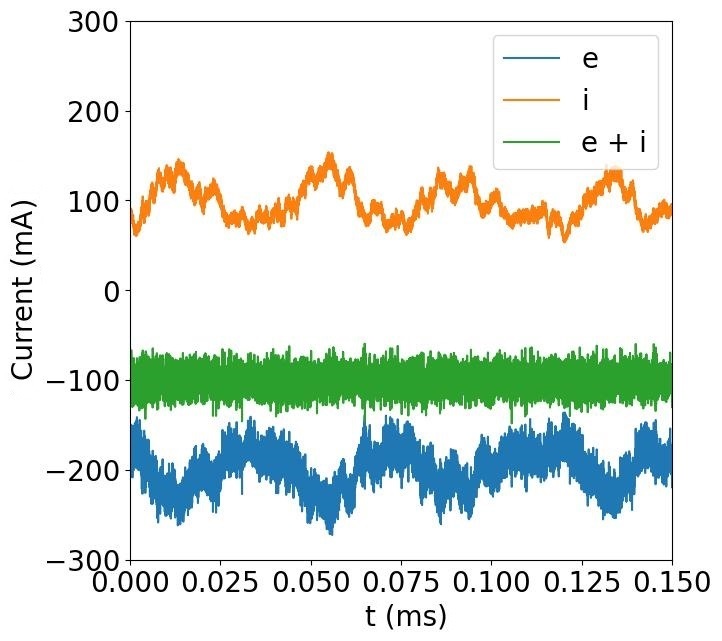}}
\captionsetup{justification=raggedright,singlelinecheck=false}
\begin{picture}(0,0)
    \put(-420,400){\textbf{(a)}} 
    \put(-420,180){\textbf{(b)}} 
    \put(-180,180){\textbf{(c)}}
    \put(-190,400){\textbf{(d)}}
\end{picture}
\vspace{-25pt}
\caption{The total electron and ion currents at the boundary as a function of time for the square geometry and circular geometries: (a) square geometry with collisions and ionization, (b) square geometry with external plasma sources (in absence of collisions and ionization); (c)   circular geometry with collisions and ionization,   $1\;$ mA injection; (d) square geometry with collisions and ionization,  $100\;$ mA injection. Note the absence of the oscillations for circular geometry, in all cases  $B = 150 \;$G. }
\label{fig:e_i_currents}
\end{figure}

Despite these similarities, the geometry of the boundary introduces an essential difference in the electron and ion flows to the wall. Most easily, this difference can be seen in the temporal behavior of the total electrons and ion currents measured at the walls. 
Figures \ref{fig:e_i_currents}  show that the square boundary modulates the total electron and ion currents at the walls at a frequency roughly four times the spoke frequency.
Note that the sum of the electron and ion current does not show these oscillations and is equal to the total injected current of $1\;$ mA.
It is worth noting that in both cases, with $1\;$ mA and $100\;$ mA electron beam injection, the value of the ion current due to the ionization is roughly equal to the electron current induced by the ionization. Consequently, the electron current to the wall is roughly double the injected current.

 The oscillations at the fourth harmonic of the spoke frequency are not related to ionization:  Figs. \ref{fig:e_i_currents} (a) and (b) exhibit similar oscillatory behavior, despite (a) being fully collisional and (b) collisionless and without ionization.
Fig. \ref{fig:e_i_currents} (c) shows that these geometry-induced oscillations disappear in the circular geometry. It is a result of geometrical modulations of particle losses as the spoke rotates against the square boundary. The modulations are also evident in the total particle inventories,  as seen in Fig. \ref{fig:CircleVsSquare-fsPartcileEvol}.

\begin{figure}[htp]
 \centering
 \captionsetup[subfigure]{skip=-20pt}
 \subcaptionbox{\label{fig:}}{\includegraphics[width=1\linewidth]{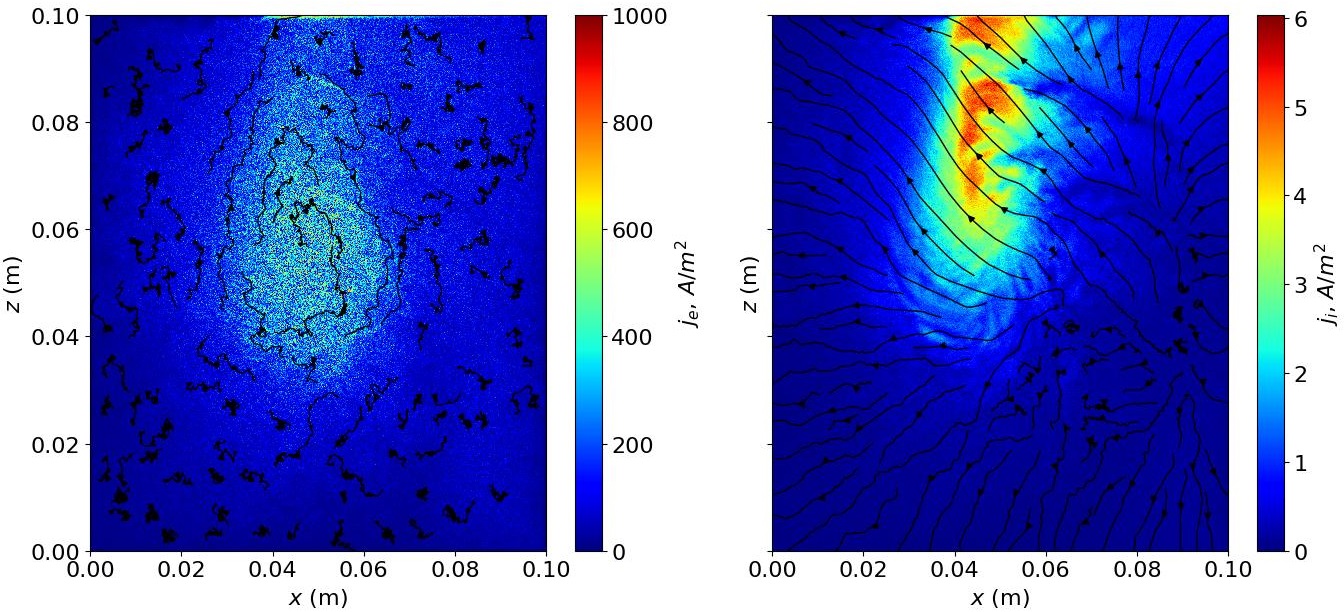}}
 \subcaptionbox{\label{}}{\includegraphics[width=1\linewidth]{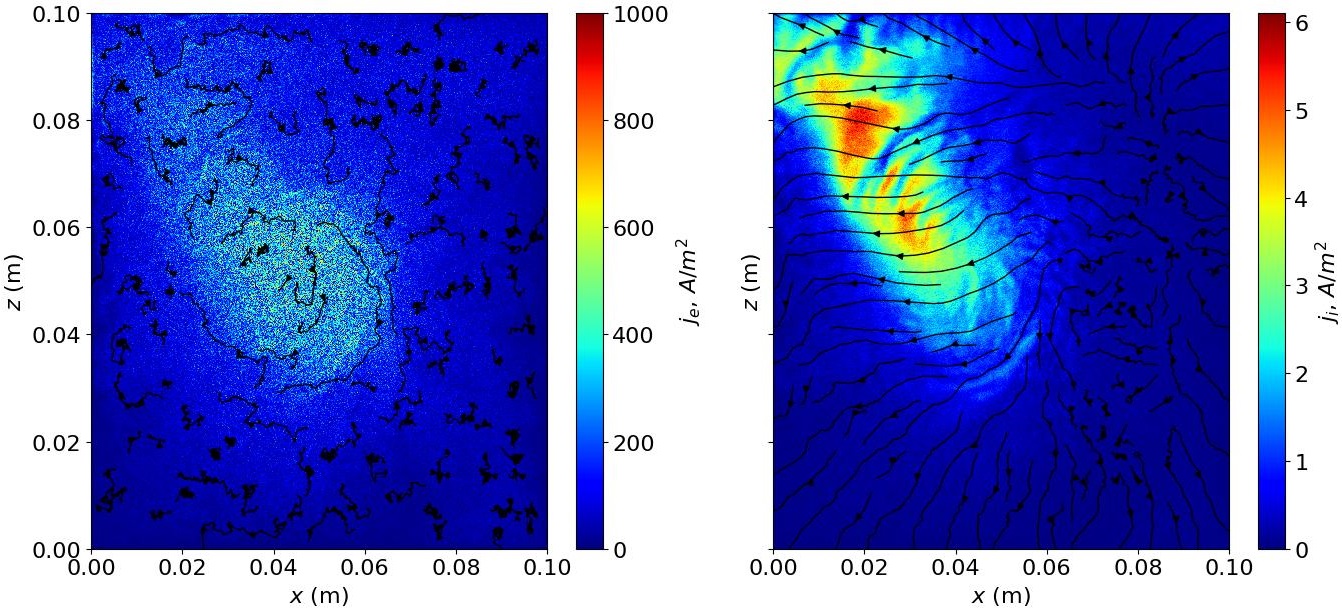}}
\captionsetup{justification=raggedright,singlelinecheck=false}
\vspace{-30pt}
\caption{Electron and ion current densities for two time frames of a spoke in the 100 mA regime.}
\label{fig:current_position}
\end{figure}

 The  Fig. \ref{fig:e_i_currents} (e)
    shows the square boundary-induced electron and ion oscillations for the case of $100\;$ mA injections.  
  In Fig. \ref{fig:current_position}, we present the current density behavior for the 100 mA case at two different time frames: one at the moment when the spoke is perpendicular to the wall and the other -- directed to the corner. The spoke-induced ion flow to the wall has both azimuthal and radial structure -- note the radial propagating ion flow blobs propagating along the spoke.   One can also note that the direction of the ion current is consistently deflected from the radial spoke direction as a result of the ion inertial forces.

% a for square geometry are similar to oscillations in the current in Fig.\ref{fig:e_i_currents} (left). \textcolor{red}{The fluctuation can be understood as follows: when the spoke is perpendicular to the wall, as shown in Fig. \ref{fig:CircleVsSquare}, the behavior is almost identical in both geometries. In circular geometry, the volume the spoke occupies remains constant as it rotates. However, in square geometry, as the spoke moves beyond the circular boundary, the volume it can occupy increases, allowing it to stretch further outward. This leads to a delay in the current flow to the wall. When the spoke is positioned diagonally, each individual current reaches a minimum, while the averaged system concentration reaches a maximum. Moving from the corner back to a perpendicular position relative to the boundary, the additional concentration and current are then deposited on the wall. This way one rotation of the spoke is going through 4 corners creating $m=4$ harmonic of a spoke.} 

% \begin{figure}[htp]
% \centering
% \captionsetup{justification=raggedright,singlelinecheck=false}
% \includegraphics[width=1\textwidth]{Mikhail_s_images/1_mA_circle/conc_1mA_circle.jpeg}
% \caption{Electron and ion concentration for circular geometry}
% \label{fig:1}
% \end{figure}

%-----------------------------------------------
%---------------- SUMMARY ----------------------
%-----------------------------------------------

\section{Summary}\label{sec:SummaryDiscussion}

In this paper, we have investigated the transitions between different types of azimuthal structures excited in $\mathbf{E}\cross \mathbf{B}$ plasma self-consistently supported by the ionization from an axial electron beam. We show that multiple $m$ spiral arm structures are excited in plasmas with zero (or slightly positive) radial electric field. With the increase of the magnetic field, electron confinement improves and a negative (inward) electric field is generated. At this point, the $m=1$ spoke mode is excited in the discharge.

We suggest here that the excitations of the spiral arms are due to the instability caused by the combination of the density gradient and ion radial flow when the condition $\mathbf{v}_{i0}\cdot \nabla n<0$ is satisfied as discussed in Ref. \onlinecite{SmolyakovPPCF2017}.
Such instability generates modes with simultaneous propagation in the radial and azimuthal directions as is observed for spiral arms. 
As in Ref. \onlinecite{TyushevPoP2023}, the excitation of the $m=1$ spoke is explained as a result of the collisionless Simon-Hoh instability due to the radial electric field. This mechanism is related to the combined effect of the density gradient and the electron $\mathbf{E}\times \mathbf{B}$ drift due to the equilibrium radial electric field when $\mathbf{E}\cdot \nabla n>0$. 
One has to note that the observed $m=1$ spoke is not the most unstable linear mode. 
The linear theory predicts that the growth rate increases with $m$ \cite{SakawaPFB1993,SmolyakovPPCF2017,BoeufPoP2023,AggarwalPoP2023}. For the higher $m$ modes the effects of the electron inertia become important and the basic instability mechanisms due to the density gradient, the equilibrium electron and ion flows are modified, and the mode frequency shifts toward the lower-hybrid range. The dissipation due to the electron collisions may also provide additional destabilization \cite{SmolyakovPPCF2017,LuckenPoP2019}. Direct excitations and interactions of small and large-scale modes were studied experimentally \cite{PrzybockiPRL2024} and in simulations\cite{BoeufPoP2019,LiangPSST2021}.
Thus the $m=1$ spoke mode should be viewed 
as a strongly nonlinear perturbation formed by the nonlinear interactions (inverse cascade) from the linear unstable high $m$ modes.

In our simulations, we directly observe high-frequency small-scale modes coexisting with the $m=1$ spoke. The high-frequency modes are observed in MUSIC and FFT  spectra in density and electric field fluctuations. Another result observed in our simulations is the heating of cold ions generated by the ionization. Such heating, as is shown in Fig. 16, which increases the ion temperature to the range of $\approx 1$ eV, is interpreted as a result of the lower-hybrid type modes. The ion heating is more pronounced in the simulations with external sources and the absence of ionization, comparing the ion temperature in the top and bottom rows (right column) of  Fig. 16. Such ion heating has been observed in $\mathbf{E}\times \mathbf{B}$ plasma experiments \cite{ChopraJAP2024}.  
 The reduced heating in the simulations with ionization can be explained as a result of the collisional dissipation. We recall that in the simulations without ionization, the energy of the injected ions is  $T_i=0.1$ while the ion energy observed in the region outside of the spoke reaches $T_i\approx 1$ eV.

Our simulations without ionization further confirm the conclusions made in Ref.\onlinecite{PowisPoP2018} that the ionization is not essential for the spoke excitation. We have performed simulations replacing the ionization with external sources that produce electrons and ions at the same rates and energies as in the self-consistent simulations with ionization. The observed patterns of the spoke and high-frequency small-scale modes remain similar suggesting that in these conditions gradient-drift instabilities, such as Simon-Hoh, are the primary sources of the observed structures.   We have  to note, however, that in some conditions the situation could be more complex. The spoke structures observed  in direct current magnetron sputtering discharges are directly identified as moving ionization zones with enhanced emissivity where the  ionization seems to be important\cite{PanjanJAP2017,GallianPSST2013}.   The difference with the modes considered here potentially may be related to the presence (in the magnetrons) of a strong electric field applied externally, as well as to the effects of the inhomogeneous magnetic field\cite{BoeufPoP2020}, where the ionization due to the electron heating provides an additional drive for the instability.    

We demonstrate here the effect of the geometry of the external boundary on the electron and ion fluxes. It is shown that the square boundary modulates the electron and ion fluxes, at the frequency four times that of the spoke rotation. However, these modulations are not related to the modulation of the ionization so the total plasma source due to ionization is not modulated. It is worth noting that the probe measurements may detect modulation of the ion and electron fluxes by non-symmetric boundaries, which may be misinterpreted as a different (from the spoke) mode. Note that for the general non-circular geometry of the boundary,   the frequency of the modulated electron and ion fluxes is not necessarily an integer multiple of the spoke frequency as in our case of the square boundary.    

% \section{Supplementary Material}
% To further illustrate the results presented in Fig. \ref{fig:DensSnap} (b-d), supplementary videos have been provided and can be accessed through the following YouTube links, Fig. \ref{fig:DensSnap-50g}: \url{ https://youtu.be/AyuNx3beVps}, Fig. \ref{fig:DensSnap-100g}: \url{https://youtu.be/5YpffBnXupU}, Fig. \ref{fig:DensSnap-220g}:\url{https://youtu.be/RlQdVlMbHGI}.
% These videos offer a dynamic view of the penning discharge with different magnetic field intensities with a square cross-section.

\acknowledgments{This work is supported in part by NSERC Canada,  SciNet-SOSCIP and the Digital Research Alliance of Canada, https://alliancecan.ca/en, and the Plasma Collaborative Research Facility at Princeton Plasma Physics Laboratory. This research used the open-source particle-in-cell code WarpX https://github.com/ECP-WarpX/WarpX, primarily funded by the US DOE Exascale Computing Project. Primary WarpX contributors are with LBNL, LLNL, CEA-LIDYL, SLAC, DESY, CERN, and TAE Technologies. We acknowledge all WarpX contributors. WarpX was run on the Mist server of SciNet-SOSCIP, https://docs.scinet.utoronto.ca/index.php/Mist.}
% The authors acknowledge with gratitude the important  discussions with ...}
 
\section*{Data availability}
The data that support the findings of this study are available from the corresponding author upon reasonable request.

\bibliographystyle{unsrt}
\bibliography{ref}
\end{document}